\documentclass{article}

\usepackage{PRIMEarxiv}

\usepackage{longtable}
\usepackage{rotating}
\usepackage{lscape}
\usepackage{float}
\usepackage{caption}
\usepackage{multirow}
\usepackage[utf8]{inputenc} 
\usepackage[T1]{fontenc}    
\usepackage{hyperref}       
\usepackage{url}            
\usepackage{booktabs}       
\usepackage{amsfonts}       
\usepackage{nicefrac}       
\usepackage{microtype}      
\usepackage{lipsum}
\usepackage{fancyhdr}       
\usepackage{graphicx}       

\title{Designing Explainable Predictive Machine Learning Artifacts: Methodology and Practical Demonstration
}

\author{
  Giacomo Welsch, Peter Kowalczyk \\
  Chair of Information Systems Engineering \\
  University of Würzburg \\
  Würzburg\\
  \texttt{\{giacomo.welsch, peter.kowalczyk\}@uni-wuerzburg.de} \\
}

\begin{document}
\maketitle

\begin{abstract}
Prediction-oriented machine learning is becoming increasingly valuable to organizations, as it may drive applications in crucial business areas. However, decision-makers from companies across various industries are still largely reluctant to employ applications based on modern machine learning algorithms. We ascribe this issue to the widely held view on advanced machine learning algorithms as "black boxes" whose complexity does not allow for uncovering the factors that drive the output of a corresponding system. To contribute to overcome this adoption barrier, we argue that research in information systems should devote more attention to the design of prototypical prediction-oriented machine learning applications (i.e., artifacts) whose predictions can be explained to human decision-makers. However, despite the recent emergence of a variety of tools that facilitate the development of such artifacts, there has so far been little research on their development. We attribute this research gap to the lack of methodological guidance to support the creation of these artifacts. For this reason, we develop a methodology which unifies methodological knowledge from design science research and predictive analytics with state-of-the-art approaches to explainable artificial intelligence. Moreover, we showcase the methodology using the example of price prediction in the sharing economy (i.e., on Airbnb).
\end{abstract}

\keywords{Design Science \and Machine Learning \and Explainable AI \and Predictive Analytics}

\section{Introduction}

Machine learning (ML) is a focal element of digitization that affects many areas of modern society: besides driving a plethora of physical and virtual products already woven into our daily lives, such as smartphones and social media platforms, ML techniques can be leveraged to power a wide range of business applications \cite{Chui2018, LeCun2015}. Although ML as an umbrella term comprises various techniques, some of which are aimed at different purposes, most ML algorithms are designed to calculate empirical predictions based on given data \cite{LeCun2015}. This prediction-oriented approach to ML is widely referred to as \textit{supervised learning}, \textit{predictive analytics}, or \textit{predictive modeling}, and initially requires at least two data sets: one for model \textit{training} and one for \textit{testing} \cite{LeCun2015, Shmueli2011}. While the former allows a given ML algorithm to "learn" patterns that connect the model input and output, the latter serves to evaluate the predictive accuracy of a trained model. In practice, if a corresponding ML model is attributed to possess a sufficient degree of predictive power, it may be deployed in a productive environment to compute real-world predictions, e.g., to support managerial decision making. The application of supervised learning in business contexts is highly relevant as it may drive applications in the fields of predictive maintenance, financial fraud detection, personalized product recommendation, and more. Consequently, the global ML market size was valued at US\$ 34.56 billion in 2021 and is expected to grow to US\$ 74.99 billion by 2028 at a compound annual growth rate of 25.7\% \cite{GrandViewResearchInc.2022}.

Given the enormous business potential of ML, a considerable number of companies have already begun to launch data analytics initiatives to automate their processes or support their decision making over the last years. Yet, Chui et al. (2018) \cite{Chui2018} report that most of these companies could (i) increase their performance or (ii) generate additional insights if they were employing applications based on more advanced ML algorithms, such as artificial neural networks. Moreover, for many decision makers from companies across various industries, the application of modern ML techniques is still perceived more as a hype than reality; just as concerning is that the majority of managers in leading positions are currently not feeling prepared for the adoption of ML \cite{Krishna2020}. We ascribe these issues particularly to one widely held view on advanced ML algorithms, namely that models generated by such algorithms are broadly regarded as “black boxes” whose complexity does not allow for uncovering associations between the model input and output \cite{Breiman2001}. Even beyond business cases that require such model interpretability (e.g., if predictions need to be explained to prudential regulators), and regardless of its predictive accuracy, human decision-makers tend to feel uncertain relying on a decision support system if they do not understand the factors that drive the system output \cite{Kayande2009}. Thus, we contend that the sustained perception of ML models as black boxes by decision makers prevents ML from diffusion, which bears the risk of entailing severe competitive consequences for the companies concerned.

We argue that academic research can and should contribute to overcome this barrier to adoption by providing (i) tools to support ML model interpretation and (ii) prototypical applications (i.e., artifacts) incorporating such tools. As for the former, we recognize that research in computer science has by now developed a range of such tools. Among these tools, one is particularly compelling as it is capable of efficiently and reliably calculating and visualizing the impact of independent input variables on an ML model output: SHAP (“SHapley Additive exPlanations”). The tool was recently presented within the scope of the seminal \textit{Nature Machine Intelligence} article by Lundberg et al. (2020) \cite{Lundberg2020}. Moreover, for the development of corresponding artifacts, we consider one paradigm in information systems (IS) to be especially suitable, as its research constantly strives to extend the boundaries of organizational capabilities: design science research (DSR) \cite{Hevner2004}. Artifacts developed in this field commonly seek to provide a novel degree of utility with respect to a given problem space and, ideally, yield insights enabled through their application. However, as we show in the literature review of this paper, there has been little research on the development of prediction-oriented artifacts leveraging approaches to black box model interpretation in DSR so far. 

We attribute this issue to the lack of methodological guidance on how to develop these artifacts; although research efforts in DSR led to several methodologies that seek to guide artifact development, none of them were deliberately designed to create artifacts powered by ML \cite{Venable2017}. Hence, current DSR methodologies do not provide a solution to the “black box problem”, i.e., they do not include a step that explicitly considers ML model interpretation. Aside from the DSR methodologies, IS research also resulted in valuable data analytics guidelines that are applicable to conducting prediction-oriented studies \cite{Fayyad1996, Muller2016,Shmueli2011}. Although these guidelines may be transferable to developing artifacts in the sense of DSR, the corresponding authors neither bridge the gap between predictive modeling and DSR nor do they provide distinct guidance on \textit{how} to efficiently uncover reliable associations underlying the output of complex ML models in general.

In the present paper we address this research gap by developing and showcasing a methodology tailored to build and interpret prediction-oriented ML artifacts. For this purpose, we build upon suitable methodologies from DSR and predictive analytics. Moreover, we consider the use of the above-mentioned SHAP. Thus, the contribution of our study is threefold. First, we contribute to the body of methodologies that guide IS research efforts by combining methodological knowledge from the complementary fields of DSR and predictive analytics with techniques that support ML model interpretation. Second, by providing a methodology to develop interpretable predictive ML artifacts, we contribute to overcome the adoption barrier of ML in organizations, as such artifacts may be explained to human decision makers. Third, from a scientific viewpoint, insights gained through prediction-oriented ML model interpretation can be used to stimulate theorizing, which may lead to new or refined theory \cite{Shmueli2011}. Artifacts developed by the use of our methodology are thus highly suited for providing additional scientific knowledge through their application, which is, according to Hevner (2007) \cite{Hevner2004}, one focal purpose of DSR.

\section{Theoretical Background and Preliminaries}
\label{sec:2}

\subsection{Design Science Research}
\subsubsection{About the Paradigm}

\begin{figure}[H]
  \centering
  \includegraphics[width=1.0\textwidth]{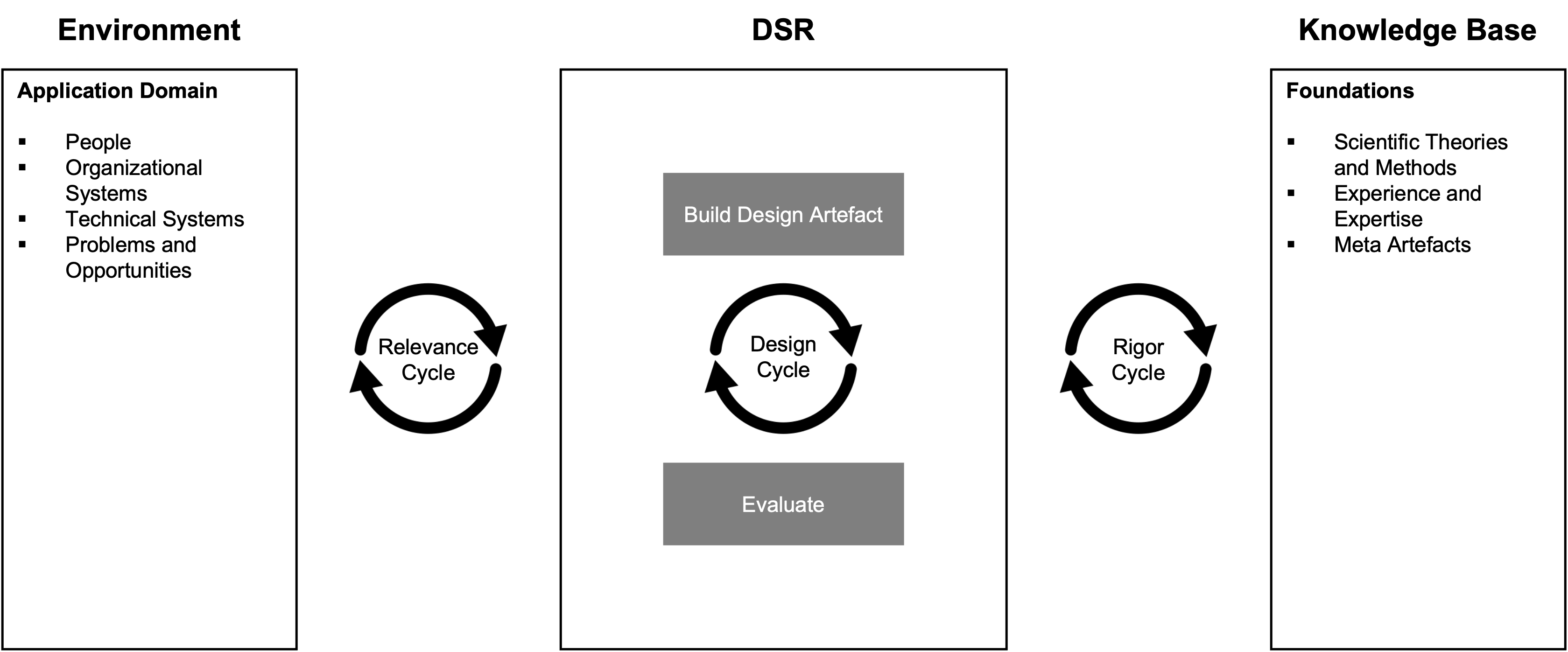}
  \caption{Three cycle view of DSR (Hevner, 2007)}
  \label{fig:fig1}
\end{figure}

Apart from behavioral research, DSR is frequently referred to as one of the two pillars of IS research \cite{Hevner2004}. While the former is particularly concerned with gaining knowledge about human-technology interaction, the latter is focused on developing useful artifacts—often in business contexts. Since artifacts developed in the course of DSR studies are—as opposed to theoretical insights gained from behavioral studies—traditionally more practice-oriented, scholars in this field first struggled to position their artifacts as scientific outcomes \cite{Peffers2018}. To overcome this hurdle to justification, research efforts led to seminal contributions which illustrate the importance of DSR as a scientific discipline \cite{Gregor2013,Gregor2007,Hevner2004,March1995}. One oft-cited framework in this context is the \textit{three cycle view} of DSR put forward by Hevner (2007) \cite{Hevner2007}. As depicted in Figure \ref{fig:fig1}, the corresponding three cycles refer to the interactions between (i) a given environment and a DSR project (\textit{relevance cycle}), (ii) the IS knowledge base and the DSR project (\textit{rigor cycle}), and, finally, (iii) the iterative process of building and evaluating a respective artifact (\textit{design cycle}). Apart from the obvious purpose of the latter, the first cycle is aimed at assuring that the artifact under construction possesses a sufficient degree of practical utility, whereas the second cycle is concerned with the artifact building upon existing knowledge to ensure its innovation. Moreover, the rigor cycle seeks a DSR project to result in valuable additions to the knowledge base, which are gained throughout artifact development and evaluation, such as theoretical insights, methodological improvements, or new meta-artifacts \cite{Hevner2007}. Notably, the three cycle view has recently been extended by a fourth cycle (\textit{change and impact cycle}) that accounts for dynamic changes in the environment into which a DSR project is embedded \cite{Drechsler2016}. Although this extension provides great value to the conceptualization of DSR, we refrain from describing it in detail, as this would go beyond the scope of our paper. However, we encourage practitioners and researchers to consider the change and impact cycle when executing a DSR project.

Besides artifact development, DSR studies often focus on the creation of design theory, i.e., prescriptive knowledge about \textit{why} and \textit{how} to build artifacts \cite{Baskerville2010,Gregor2013,Gregor2007}. As the boundaries between these outcome-types tend to blur, some DSR contributions ought to position themselves as hybrids providing some degree of prescriptive knowledge and innovative practical utility. Therefore, in order to shed light on the spectrum between artifact and theory and, thus, DSR contributions, Baskerville, Baiyere, Gregor, Hevner, \& Rossi (2018) \cite{Baskerville2018} propose five proper DSR objectives. Those are (i) \textit{technology and science evolutions}, (ii) \textit{design artifacts}, (iii) \textit{design theories}, (iv) \textit{DSR processes}, and (v) \textit{DSR impacts}. While the (i) first objective ensures that DSR efforts clarify their contribution to science-technology evolution, (ii) artifacts are most often designed before the development of corresponding theories and must be evaluated in a DSR project. (iii) Design theories, on the other hand, represent design knowledge and should be recognized as inputs and outputs of a DSR endeavor. Furthermore, (iv) DSR processes serve to conduct DSR studies properly, whereas (v) DSR impacts ensure that respective research efforts impact both practice and research, which is in line with the three cycle view.

Against the backdrop of the DSR objectives, we seek to develop a DSR process (i.e., a methodology) within the scope of this article. This process aims to provide guidance on how to develop and interpret ML-driven artifacts.

\subsubsection{Design Science Research Methodologies}

\begin{figure}[H]
  \centering
  \includegraphics[width=1.0\textwidth]{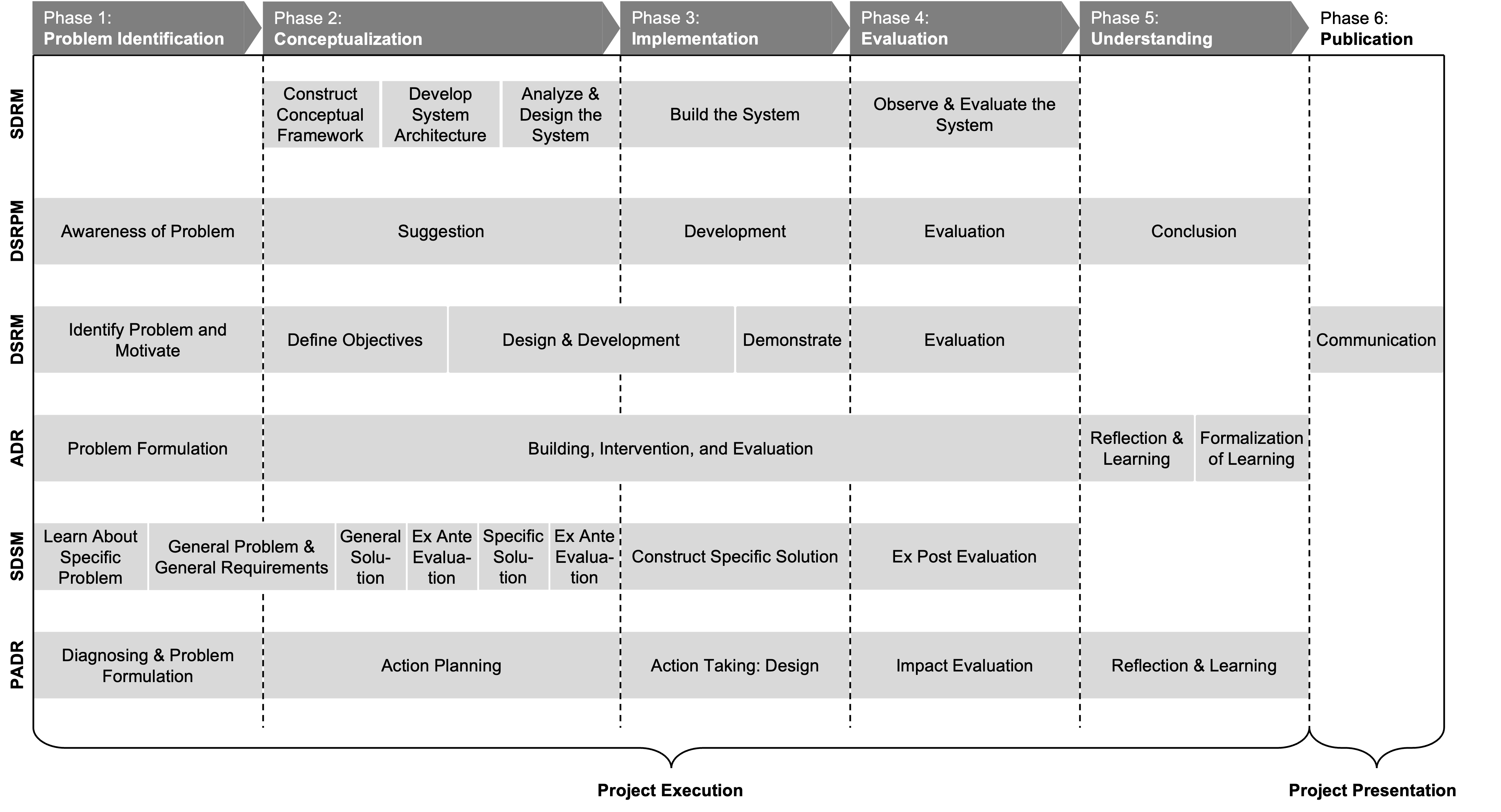}
  \caption{DSR methodologies}
  \label{fig:fig2}
\end{figure}

To provide guidance on how to conduct DSR studies, scholars developed several methodologies over the years. Venable et al. (2017) \cite{Venable2017} summarize the six most commonly used methodologies in their article. These are (i) \textit{systems development research methodology} (SDRM) \cite{Nunamaker1990}¸ (ii) \textit{DSR process model} (DSRPM) \cite{Vaishnavi2015, Vaishnavi2004,Vaishnavi2008}, (iii) \textit{design science research methodology} (DSRM) \cite{Peffers2007}, (iv) \textit{action design research} (ADR) \cite{Sein2011}, (v) \textit{soft design science methodology} (SDSM) \cite{Baskerville2009}, and (vi) \textit{participatory action design research} (PADR) \cite{Bilandzic2011}. They are depicted in Figure \ref{fig:fig2}. All of them are particularly suitable for artifact development but are also applicable for creating design theories.

After examining the methodologies, we were able to derive six generic consecutive phases of a DSR project, as shown in the figure. All the steps of each methodology are assignable to these phases, as they reflect the progress of a DSR endeavor. The phases are (i) \textit{problem identification}, (ii) \textit{conceptualization}, (iii) \textit{implementation}, (iv) \textit{evaluation}, (v) \textit{understanding}, and, finally, (vi) \textit{publication}. While the first five steps constitute the execution of a DSR project, the final step is concerned with making the study available to an audience that consists of researchers and practitioners, typically in the form of a scholarly publication. Although Figure \ref{fig:fig2} suggests the conceptualization phase to be longer than the other phases, the extent of the different phases is not by any means determined and may vary across projects. The actual reason of the conceptualization phase being illustrated by a longer arrow is that particularly SDSM, but also SDRM, put a stronger emphasis on conceptualization, i.e., these methodologies describe conceptualization extensively by using numerous steps. Moreover, the width of the single steps in the figure does not inherently imply the extent of each step—it is merely due to the correct assignment of steps and phases.

Observing Figure \ref{fig:fig2} further, we recognize that not all methodologies consider all the phases derived. However, this does not mean that the corresponding authors consider missing phases as unimportant. Instead, this means that the authors do not provide distinct guidance on how to perform these phases because their respective methodologies put more emphasis on the phases considered.

In this paper, we focus on project execution. Therefore, we only consider the first five phases of a DSR endeavor, thus neglecting the publication phase. We substantiate this decision by arguing that providing guidance on how to publish a research project would be beyond the scope of our research agenda. However, our methodology is well-suited to present a DSR project throughout a publication.

\subsection{Predictive Analytics}
\label{sec:2.2}

\subsubsection{About the Paradigm}

Predictive analytics refers to a type of statistical modeling that aims to make empirical rather than theoretical predictions \cite{Delen2018, Shmueli2011}. While the latter deductively arise from theoretical models—typically in form of a priori formulated hypotheses by a researcher—prediction-oriented models and, thus, their predictions arise inductively from data. Shmueli \& Koppius (2011) \cite{Shmueli2011} refer to the explanation-oriented, theory-driven approach to statistical modeling as \textit{explanatory modeling} and to the prediction-oriented, data-driven approach as \textit{predictive modeling}. “[W]hereas explanatory modeling seeks to minimize model bias (i.e., specification error) to obtain the most accurate representation of the underlying theoretical model, predictive modeling seeks to minimize the combination of model bias and sampling variance” \cite{Shmueli2011}. As a consequence, explanatory models are typically validated by means of goodness-of-fit tests, the R\textsuperscript{2} metric, and p-values that support hypotheses rejection and confirmation. In contrast, predictive models are typically built by applying appropriate methods (e.g., ML algorithms) to a \textit{training} data set and evaluated by calculating out-of-sample predictions using a second \textit{test} data set. Here, evaluation metrics such as the mean absolute error (MAE), mean squared error (MSE), or mean absolute percentage error (MAPE) may be utilized in case the trained model serves to solve a \textit{regression} problem (i.e., if the predicted target is a continuous variable). Moreover, if the model serves to solve a \textit{classification} problem (i.e., if the predicted target is a discrete variable), metrics such as the accuracy, precision, recall, or $F_\beta$-score may be used. While IS has a rich history in empirical explanatory research, prediction-oriented studies are traditionally rather underrepresented \cite{Shmueli2011}.

Against this background, academic researchers increasingly started to emphasize the scientific value of predictive modeling, often with reference to the rich data availability that has emerged in recent years \cite{Agarwal2014, Delen2018, Muller2016}. Aside from holding the potential to complement explanatory research efforts in behavioral research, as outlined by Shmueli \& Koppius (2011) \cite{Shmueli2011}, predictive modeling is also well-suited to drive artifacts that rely on empirical predictions, such as various types of decision support systems \cite{Delen2018}. For the training of models capable of calculating such predictions, a wide range of different methods exists. Technically, even more traditional statistical methods, such as linear or logistic regression, may be trained to predict new data in the sense of predictive modeling. The approach of using such “white box” methods to train predictive models has the obvious advantage of producing models that are easy to interpret. However, predictive modeling typically unfolds its full potential (i.e., it typically yields the most accurate predictions) in cases where complex  black box ML algorithms, such as tree ensembles or neural networks, are trained by the use of large data sets \cite{Lundberg2017a, Muller2016}. Therefore, in this paper, we focus on the development and interpretation of artifacts driven by sophisticated ML algorithms that are broadly considered black boxes.

\subsubsection{Predictive Analytics Methodologies}
\label{sec:2.2.2}

In order to guide IS research concerned with predictive modeling, several methodologies have been developed. We now present three methodologies that are oft-cited, published in leading outlets, and particularly suitable for our study.

The first is the process for \textit{knowledge discovery in databases} (KDD) put forward by Fayyad et al. (1996) \cite{Fayyad1996}. Although the process is not exclusively designed for predictive modeling, it emphasizes the broader topic of data-driven analyses (i.e., data mining) and incorporates a final step for pattern interpretation that is aimed at gaining knowledge. Therefore, its fundamental idea and, moreover, its generic steps are highly transferable to our study. However, because of recent advances in data availability and ML, the descriptions of KDDs steps—due to the maturity of the process—neglect aspects which are crucial for contemporary predictive analytics efforts, such as data partitioning or black box model interpretation.

The second methodology we consider are the \textit{steps in building an empirical model} (SBEM) proposed by Shmueli and Koppius (2011) \cite{Shmueli2011}. As opposed to KDD, SBEM is focused on building predictive empirical models. However, the authors—in their seminal article—stress the roles of predictive modeling in behavioral research rather than in DSR—which becomes particularly evident in their exemplary showcase considering a popular model from technology acceptance research. Moreover, even though the final step of the methodology (i.e., \textit{model use \& reporting}) may technically be applied to model interpretation, SBEM does not provide guidance on how to interpret predictive models.

\begin{figure}[H]
  \centering
  \includegraphics[width=1.0\textwidth]{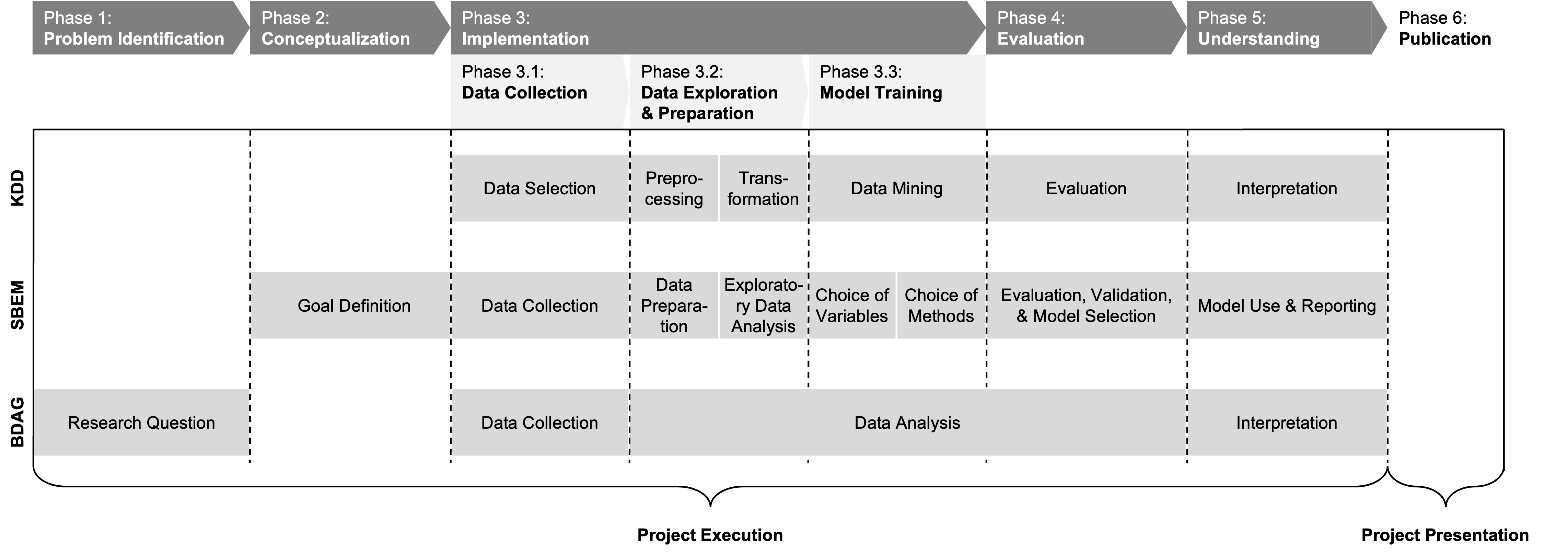}
  \caption{Predictive modeling methodologies}
  \label{fig:fig3}
\end{figure}

Finally, we consider the \textit{big data analytics guidelines} (BDAG) presented by Müller et al. (2016) \cite{Muller2016}. BDAG is predominantly focused on analyzing vast data sets (“big data”) without the compelling necessity of employing predictive modeling. However, the authors repeatedly emphasize the application of ML algorithms for prediction—therefore, we constitute BDAG to be well-suited to guide predictive modeling efforts. As opposed to SBEM, BDAG explicitly incorporates a step that considers ML model interpretation. Unfortunately, at the time the article by Müller et al. (2016) \cite{Muller2016} was written, a sophisticated tool to support the efficient and reliable interpretation of any black box ML model, such as SHAP \cite{Lundberg2020, Lundberg2017a}, was not yet available. Hence, BDAG does not provide distinct guidance on how to interpret black boxes using such a tool. Thus, our methodology does not only bridge the gap between BDAG and DSR, but also aims to extend the methodology's interpretation step by considering the use of SHAP.

In Figure \ref{fig:fig3}, we assigned the steps of the methodologies for conducting predictive studies to the phases of a DSR project, as derived from the DSR methodologies. Examining the figure, we find that only SBEM roughly incorporates conceptualization (in form of the step \textit{goal definition}), which is—according to all of the DSR methodologies—a crucial step within the scope of a DSR endeavor. Yet, the predictive analytics methodologies put more emphasis on implementation than on the conceptualization phase. After examining corresponding articles about the methodologies, we were able to derive three consistent sub-phases the implementation phase can be divided into if the goal is to build a predictive model. These are (i) \textit{data collection}, (ii) \textit{data exploration and preparation}, and (iii) \textit{model training}. The corresponding steps of each methodology can be assigned to these sub-phases, as shown in Figure \ref{fig:fig3}. As for the widths of the steps in the figure we—again—note that these are due to the correct assignment of steps and phases, i.e., the width does not necessarily correspond to the extent of a step. Moreover, in line with our decision to neglect phase 6 in our methodology, none of the methodologies provide guidance on how to write and publish a research article.

In the remainder of this paper, we build upon the insights gained from examining the DSR and predictive modeling methodologies.

\subsection{Prediction-Oriented ML Artifacts and Black Box Interpretation}

\subsubsection{Value for IS Research}
\label{sec:2.3.1}

In this section, we shed light on the value of artifacts driven by black box ML models and approaches that enable the interpretation of such artifacts for IS research. Notably, despite the debates about the differences between ML and artificial intelligence (AI) on the part of scholars as well as practitioners \cite{Iriondo2018}, such interpretation approaches are broadly referred to as \textit{explainable AI} approaches (and not \textit{explainable ML} approaches). For the sake of clarity, and in line with the deliberations of Iriondo (2018) \cite{Iriondo2018}, in this paper, we refer to ML as a branch of AI that focuses on algorithms, which aim to learn from experience represented by data. In the upcoming subsection, we briefly present a selection of explainable AI approaches suitable for ML model interpretation; for now, we note that the before-mentioned SHAP is currently the most advanced of these approaches. Therefore, we utilize SHAP as an instance of explainable AI throughout this paper. However, our methodology does not necessarily rely on using SHAP, as it is conceivable that, in the future, even more sophisticated tools emerge which have advances over SHAP. Moreover, some research projects may consider the use of other existing tools, e.g., due to project-specific requirements.

\begin{figure}[H]
  \centering
  \includegraphics[width=1.0\textwidth]{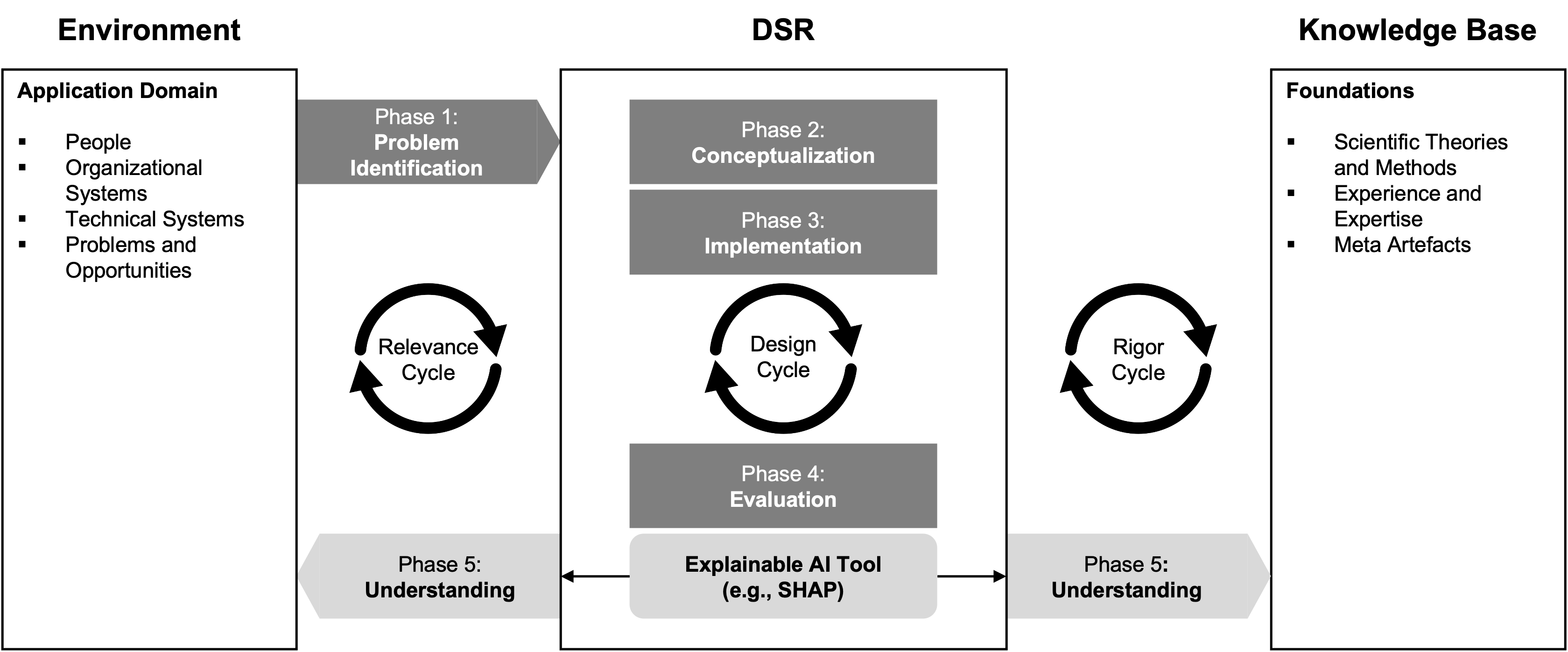}
  \caption{Value of explainable AI in DSR in case an artifact is powered by prediction-oriented ML}
  \label{fig:fig4}
\end{figure}

To highlight the value of DSR studies concerned with predictive ML modeling and explainable AI for IS research, we draw upon the before-mentioned three cycle view of DSR as shown in Figure \ref{fig:fig4}. Here, we assigned the steps of a DSR project, which we derived from the methodologies, to the corresponding cycles. We now briefly elaborate on this assignment.

First, we consider phase 1—as being the initiator of a DSR project—to be part of the relevance cycle. We substantiate this view by pointing out that DSR has its roots in solving problems that arise from the environment rather than the knowledge base. In the words of Hevner et al. (2004) \cite{Hevner2004}:

\hspace*{10mm} \textit{The environment defines the problem space \cite{Simon1996} in which reside the phenomena of interest.}

As for a prediction-oriented ML artifact, a corresponding problem might be the emergence of high costs due to repeatedly failing machines that are crucial within the manufacturing process of a company. In this subsection, we continue to consider this as an exemplary problem that serves to illustrate the purposes of the phases within the three cycle view.

Next, we assigned the phases 2, 3, and 4 to the design cycle, as these are obviously concerned with the iterative process of artifact development and evaluation as described by Hevner (2007) \cite{Hevner2007}. Despite being assigned to the design cycle, these three phases are also interwoven into the relevance and the rigor cycle, at least for problems that may be addressed by means of prediction-oriented ML artifacts. This means that these phases must consider the environment (e.g., data availability, data quality, use case) as well as the knowledge base (e.g., methods, guidelines, evaluation metrics). With regard to the exemplary problem of a manufacturing company described above, a corresponding artifact might utilize data gathered by sensors that aim to monitor the conditions of the machines (i.e., data from the environment) in order to find patterns that might predict machine failure leveraging state-of-the-art ML algorithms (i.e., methods from the knowledge base). Such artifact would need to possess a sufficient degree of predictive power, which would be assessed using techniques and metrics from the knowledge base. If such artifact would yield sound predictions, it would be capable of being deployed to avoid machine failure, e.g., by initiating preventive maintenance jobs. This would provide utility to the environment and at least experience to the knowledge base (maybe more, depending on the findings of the project).

Finally, and most importantly for our research agenda, we state that the employment of an explainable AI tool, such as SHAP, at the end of a design cycle (i.e., when the artifact is built and evaluated), enables analysts to gain insights into the factors that drive the artifact’s predictions. Should the artifact possess a sufficient degree of predictive power, such insights are especially valuable to inform (i) decision makers in organizations and (ii) researchers. Therefore, we consider phase 5 to affect both the environment and the knowledge base, and, thus, to be part of the relevance and the rigor cycle. With respect to the exemplary manufacturing company, such insights gained from ML artifact interpretation might be that the temperature of a given machine component is crucial for predicting machine failure during production. If this were the case, decision makers—aside from being more likely to adopt the artifact due to its interpretability—could consider taking actions either to prevent the machine component from reaching a critical temperature or to replace the component with a more robust substitute. Moreover, as for researchers, such insight might help to develop theories for resilient shop-floor design or to extend the set of antecedents that explain machine failure in a theoretical model in the field of production research.

However, we note that associations uncovered by using explainable AI tools are not to be confused with causal explanations, as such tools are not capable of providing the latter. Yet, combined with proper logical reasoning and/or further research, such associations are well-suited to deliver indications that point towards causality. Beyond that, even in cases where the examination of these associations does not result in uncovering novel causal explanations, the associations may still serve to enhance a given understanding of a phenomenon, e.g., by serving as specifications of the surrounding of this phenomenon.

\subsubsection{Explainable AI}

The research field of explainable AI has its roots in the oft-cited tension between accuracy and interpretability of predictive models \cite{Breiman2001, Lundberg2017a, Muller2016}: Models that yield particularly accurate predictions typically tend to be complex black boxes that are hard to interpret, whereas interpretable white box models, such as linear models, typically tend to lack predictive power. Apparently, this tension often prompts users to prefer white box models—due to their ease of interpretation—over the better performing black boxes \cite{Lundberg2017a}. In order to remedy the tradeoff between accuracy and interpretability, scholars in the field of explainable AI recently proposed approaches that enable the interpretation of models, which have long been considered black boxes, such as ensemble or deep learning models. One of these approaches is SHAP, which utilizes metrics (i.e., \textit{SHAP values}) from cooperative game theory as a reliable measure of feature importance. As mentioned above, we currently identify SHAP to be the most advanced approach to black box model interpretation—notwithstanding, we recognize that popular approaches exist that are comparable to SHAP. Therefore, following the deliberations of Lundberg \& Lee (2017) \cite{Lundberg2017a}, we briefly provide an introduction to the rationale behind these and SHAP and explain why SHAP is currently preferable over other approaches.

Lundberg \& Lee (2017) \cite{Lundberg2017a} point out that the two reasons why white box models are capable of providing explanations of the associations between the features (i.e., its input variables) and the target (i.e., its output variable) are that the model itself (i) represents these associations and (ii) is easy to understand. Although black box models also associate features with the target, they are not easy to understand, which is why they do not provide convenient explanations. Therefore, explainable AI approaches typically seek to find a relatively simple, interpretable \textit{explanation model} $g$ that approximates a given black box model $f$. For this purpose, SHAP, as well as other approaches, build upon \textit{local explanations}. Local explanations aim to explain each prediction $f(x)$ based on a single input $x$, i.e., features from one observation. In this context, explanation models typically map $x$ to simplified inputs $x'$ by using a mapping function $x=h_x (x')$. Such explanation models further seek to ensure $g(z') \approx f(h_x (z'))$ in cases where $z' \approx x'$. Here, one frequently used explanation model is the following \cite{Lundberg2017a}:

\begin{equation}
\label{equ:equ1}
g(z') = \phi_0 + \sum_{i=1}^{M} \phi_i z'_i ,
\end{equation}

where $z' \in {0,1}^M$, $M$ is the numer of simplified input features, and $\phi_i \in \mathbb{R}$.

Lundberg \& Lee (2017) \cite{Lundberg2017a} refer to methods that build upon Equation \ref{equ:equ1} as \textit{additive feature attribution methods}, which “attribute an effect $\phi_i$ to each feature, and summing the effects of all feature attributions approximates the output $f(x)$ of the original model.” Since additive feature attribution methods (i) commonly serve to drive explainable AI approaches \cite{Casalicchio2019, Lundberg2017a} and (ii) can be utilized to explain predictions both on a local (i.e., feature importance of an individual prediction) and global level (i.e., feature importance of an entire model) \cite{Lundberg2020}, we focus on such methods. Besides SHAP, popular examples of such approaches are \cite{Lundberg2017a}:

\begin{itemize}
\item \textbf{LIME} \cite{Ribeiro2016}: The \textit{local interpretable model-agnostic explanation} method (LIME) explains model predictions using local approximations of the model around the prediction at hand. It may be used to explain the predictions of any black box model.
\item \textbf{DeepLIFT} \cite{Shrikumar2017}: This recursive additive feature attribution method particularly serves to explain the predictions of deep learning models.
\item \textbf{Layer-Wise Relevance Proposition} \cite{Bach2015}: This method is a special case of DeepLIFT and also serves to explain the predictions of deep learning models.
\item \textbf{Classic Shapley Value Estimation}: There are several methods that employ Shapley values from cooperative game theory to explain model predictions: Shapley regression values \cite{Lipovetsky2001}, Shapley sampling values \cite{Strumbelj2014}, and quantitative input influence \cite{Datta2016}. While the first assigns effects of feature attributions to linear models, the remaining two methods are technically applicable to any model.
 \end{itemize}
 
In contrast to each of these approaches, SHAP unifies the whole class of additive feature attribution methods in order to propose SHAP values as a unified measure of feature importance, which are applicable to any black box model (i.e., they are \textit{model-agnostic}). As a consequence, SHAP values have several desirable properties compared to the other approaches, e.g., better alignment with human intuition, faster calculation of optimal explanations, being the only feature attribution method to provide consistent feature attributions for tree ensembles, and more \cite{Lundberg2020,Lundberg2017,Lundberg2017a}. SHAP uses the so-called Shapley values (or SHAP values) for the calculation of feature attributes $\phi_i$ as a reliable measure of the importance of features with respect to a prediction. These are now explained according to the paper by Lundberg and Lee \cite{Lundberg2017a}. 

SHAP values originate in cooperative game theory and are a measure of the importance of features within statistical models in the presence of multicollinearity. Here, $S$ is a choice of features for which $S \subseteq F$ holds, where $F$ represents all features. The objective is to assign an importance value to each feature with respect to a prediction of a model under consideration. For this purpose, for a given feature, a model $f_{SU\{i\}}$ with this feature and a model $f_{S}$ without this feature are trained. Then, the predictions of the two trained models are compared using the features of relevance: $f_{SU\{i\}}(x_{SU\{i\}})-f_{S}(x_{S})$, where $x_{S}$ are the feature values of the set $S$. Since the effect of withholding a feature depends on the values of the other features used in the model, the differences described are calculated for all possible subsets $S \subseteq F \setminus \{i\}$. Now the SHAP values can be calculated as a weighted average of all combinations of the possible differences: 

\begin{equation}
\label{equ:equ2}
\phi_{i} = \sum_{S \subseteq F \setminus \{i\}} \frac{\vert S \vert ! ( \vert F  \vert -  \vert S  \vert - 1)!}{ \vert F  \vert!} [ f_{SU\{i\}}(x_{SU\{i\}})-f_{S}(x_{S})]
\end{equation}

For these SHAP values, the previously mentioned mapping function $h_{x}$ assigns either the value one or zero to the features, where one means that a feature is considered in the model (i.e. it is important for the corresponding prediction). Zero, on the other hand, means that it is not considered (i.e., it is not important for the prediction). Moreover, $\phi_{i}= f_{\emptyset}(\emptyset)$ must hold for SHAP values to correspond to additive feature attributes. As it would go beyond the scope of this paper to mention all the merits of SHAP, we hereby refer to the referenced articles by Lundberg et al. and to the GitHub page of the corresponding Python package\footnote{\url{https://github.com/slundberg/shap}} for a deeper understanding about the tool.

In their study, Tritscher et al. (2020) \cite{tritscher2020} show that SHAP currently provides the most reliable explanations for black box models and their predictions compared to other prominent XAI approaches, closely followed by LIME. Furthermore, SHAP values have several advantages over other feature attributes. These include that (i) they can be interpreted more intuitively by humans, (ii) they can be used to compute optimal explanations more quickly, and (iii) they are the only feature attributes that provide consistent explanations for decision tree-based models (e.g., Random Forests or XGBoost) \cite{Lundberg2017,Lundberg2020,Lundberg2017a}. For these reasons, SHAP is used as a demonstrative XAI approach in the course of this paper.

\section{Literature Review}

In this section, we present the results of a literature review we conducted. These results provide an overview of research endeavors that are concerned with (i) the design of a prediction-oriented black box artifact and (ii) the explanation of its predictions.
In order to gather a representative sample of research articles, we searched five databases (i.e., Google Scholar, AIS electronic Library, IEEE Xplore, ACM Digital Library and EBSCOhost) using a query that consisted of the following five building blocks, which had to appear in the full texts of the articles: (i) “machine learning” \textit{or} “artificial intelligence”; (ii) “predict”, “predictive”, \textit{or} “prediction”; (iii) “black box”; (iv) "interpretation", "explanation", "explainable", \textit{or} "interpretable"; and (v) “design science”. This search resulted in 1501 articles (1040 in Google Scholar, 387 in the AIS electronic library, 25 in IEEE Xplore, 10 in the ACM Digital Library, and 39 in EBSCOhost)\footnote{We used the following search query for Google Scholar: ("machine learning" OR "artificial intelligence") AND ("predict" OR "predictive" OR "prediction") AND "black box" AND ("interpretation" OR "explanation" OR " interpretable" OR "explainable ") AND "design science". For the remaining databases, we adapted the query’s syntax according to their respective requirements. We conducted the search on 17\textsuperscript{th} of April 2022.}. We then manually searched these articles for (i) peer-reviewed studies employing (ii) predictive modeling and (iii) explainable AI to build prototypical applications (i.e., artifacts). Although the vast majority of the articles either present prediction-oriented studies that refrain from model interpretation or non-quantitative studies, numerous authors stress the importance of explainable AI. Yet, we only identified 35 articles that met our search criteria. Among these articles, twelve build their research upon existing methodologies while the remaining 23 do not organize their research according to a standard process (see Table \ref{tab:tab1}).

\begin{table}[H]
\caption{Methodologies used in studies that build artifacts employing predictive modeling and explainable AI}
\makebox[\linewidth]{
\begin{tabular}{|l|c|c|c|}
\hline
\textbf{Publication}                                  & \textbf{DSR methodology} & \textbf{Other methodology} & \textbf{No methodology} \\ \hline
Müller et al. (2016) \cite{Muller2016}                                 &                          & x                          &                         \\ \hline
Bohanec, Robnik-Šikonja, \& Borštnar (2017a) \cite{Bohanec2017a}         & x                        & x                          &                         \\ \hline
Bohanec, Robnik-Šikonja, \& Borštnar (2017b) \cite{Bohanec2017}         & x                        & x                          &                         \\ \hline
Ming, Qu, \& Bertini (2018) \cite{Ming2018}                          & \textbf{}                & \textbf{}                  & x                       \\ \hline
Ventura, Cerquitelli, \& Giacalone (2018) \cite{Ventura2018}            & \textbf{}                & \textbf{}                  & x                       \\ \hline
Colace et al. (2019) \cite{Colace2019}                                 &                          &                            & x                       \\ \hline
Eitle \& Buxmann (2019) \cite{Eitle2019}                              & x                        &                            &                         \\ \hline
Yeganejou, Dick, \& Miller (2019) \cite{Yeganejou2019}                    & \textbf{}                & \textbf{}                  & x                       \\ \hline
Alvanpour, Das, Robinson, Nasraoui, \& Popa (2020) \cite{Alvanpour2020}   &                          & x                          &                         \\ \hline
Bramhall, Horn, Tieu, \& Lohia (2020) \cite{Bramhall2020}                &                          &                            & x                       \\ \hline
Dolk, Kridel, Dineen, \& Castillo (2020) \cite{Dolk2020}             &                          &                            & x                       \\ \hline
Harl, Weinzierl, Stierle, \& Matzner (2020) \cite{Harl2020}          & \textbf{}                & \textbf{}                  & x                       \\ \hline
Kim, Srinivasan, \& Ram (2020) \cite{Kim2020}                       & \textbf{}                & \textbf{}                  & x                       \\ \hline
Mehdiyev \& Fettke (2020) \cite{Mehdiyev2020}                            & x                        & \textbf{}                  & \textbf{}               \\ \hline
Yadam, Moharir, \& Srivastava (2020) \cite{Yadam2020}                 & \textbf{}                & \textbf{}                  & x                       \\ \hline
Zhang, Du, \& Zhang (2020) \cite{Zhang2020}                           &                          &                            &                         \\ \hline
Asmussen, Jørgensen, \& Møller (2021) \cite{Asmussen2021}                & x                        & x                          &                         \\ \hline
Barfar, Padmanabhan, \& Hevner (2021) \cite{Barfar2021}                &                          &                            & x                       \\ \hline
Garriga, Aarns, Tsigkanos, Tamburri, \& Heuvel (2021) \cite{Garriga2021} & x                        &                            &                         \\ \hline
Han et al. (2021) \cite{Han2021}                                    &                          &                            & x                       \\ \hline
Herm, Wanner, Seubert, \& Janiesch (2021) \cite{Herm2021}            &                          & x                          &                         \\ \hline
Johnson, Albizri, Harfouche, \& Tutun (2021) \cite{Johnson2021}         &                          &                            & x                       \\ \hline
Liu, Du, Hong, \& Fan (2021) \cite{Liu2021}                         &                          &                            & x                       \\ \hline
Mehdiyev \& Fettke (2021) \cite{Mehdiyev2021}                            & x                        &                            &                         \\ \hline
Mombini et al. (2021) \cite{Mombini2021}                                &                          &                            & x                       \\ \hline
Pereira et al. (2021) \cite{Pereira2021}                                &                          &                            & x                       \\ \hline
Velichety \& Ram (2021) \cite{Velichety2021}                              &                          &                            & x                       \\ \hline
Wang et al. (2021) \cite{Wang2021}                                   &                          &                            & x                       \\ \hline
Wastensteiner, Michael Weiss, Haag, \& Hopf (2021) \cite{Wastensteiner2021}   & x                        &                            &                         \\ \hline
Zhou, Wang, Ren, \& Chen (2021) \cite{Zhou2021}                      &                          &                            & x                       \\ \hline
Bodendorf, Xie, Merkl, \& Franke (2022) \cite{Bodendorf2022}              & x                        &                            &                         \\ \hline
Johnson, Albizri, Harfouche, \& Fosso-Wamba (2022) \cite{Johnson2022}   &                          &                            & x                       \\ \hline
Kowalczyk, Röder, Dürr, \& Thiesse (2022) \cite{Kowalczyk2022}            &                          &                            & x                       \\ \hline
Tao, Zhou, \& Hickey (2022) \cite{Tao2022}                          &                          &                            & x                       \\ \hline
Yang, Yuan, \& Lau (2022) \cite{Yang2022}                            &                          &                            & x                       \\ \hline
\end{tabular}
}
\label{tab:tab1}
\end{table}

These articles present studies in the areas of business process management \cite{Asmussen2021,Bodendorf2022,Garriga2021,Harl2020,Johnson2022,Mehdiyev2020,Mehdiyev2021}, healthcare \cite{Han2021,Johnson2021,Kim2016,Ming2018,Mombini2021}, human resource management \cite{Bramhall2020,Colace2019}, image recognition \cite{Alvanpour2020,Ventura2018,Yeganejou2019}, sales \cite{Bohanec2017,Bohanec2017a,Eitle2019}, security \cite{Kowalczyk2022,Tao2022,Wang2021,Yang2022,Zhang2020,Zhou2021} and others \cite{Barfar2021,Dolk2020,Herm2021,Liu2021,Muller2016,Pereira2021,Velichety2021,Wastensteiner2021,Yadam2020}. Moreover, all of these articles were published in recent years, with 27 of them being published since 2020. Thus, we consider their authors pioneers in the development of interpretable predictive ML artifacts and further point out that their recent publication highlights the topicality of explainable AI in predictive modeling. On the downside, the examination of the articles unveils the lack of a consistent methodology that guides their underlying research efforts. This becomes particularly evident in the articles by Bohanec et al. (2017) \cite{Bohanec2017,Bohanec2017a} and Asmussen et al. (2021) \cite{Asmussen2021}. In these articles, the authors are compelled to combine a DSR methodology (i.e., ADR) with a non-academic methodology for data mining (i.e., CRISP-DM) \cite{Wirth2000} to train and interpret ML models. Moreover, the six studies of Eitle and Buxmann (2019) \cite{Eitle2019}, Mehdiyev and Fettke (2020, 2021) \cite{Mehdiyev2020,Mehdiyev2021}, Garriga et al. (2021) \cite{Garriga2021}, Wastensteiner et al. (2021) \cite{Wastensteiner2021} and Bodendorf et al. (2022) \cite{Bodendorf2022} adapted methodologies from DSR to their settings, even though these methodologies do not natively account for all of the steps required to build predictive models. Although the authors—due to their experience—broadly managed to evade the pitfalls that may come with such adaption, we argue that the presence of a suitable methodology for the development of interpretable predictive black box artifacts may not only contribute to prevent methodological mistakes in the future but may even encourage more scholars to conduct research in this area. Despite the described articles, our literature review also yielded the article by Müller et al. (2016) \cite{Muller2016}, which we already discussed in section \ref{sec:2.2.2}, as it presents and showcases BDAG. The remaining studies do not follow a standard methodology. As a consequence, the studies are highly heterogeneous, which hampers comparing their results. Although we recognize that, due to the variety across studies, some of them are hardly comparable, we note that the widespread use of a standard methodology would particularly foster the comparability of similar studies.

\section{Methodology}

\begin{figure}[H]
  \centering
  \includegraphics[width=1.0\textwidth]{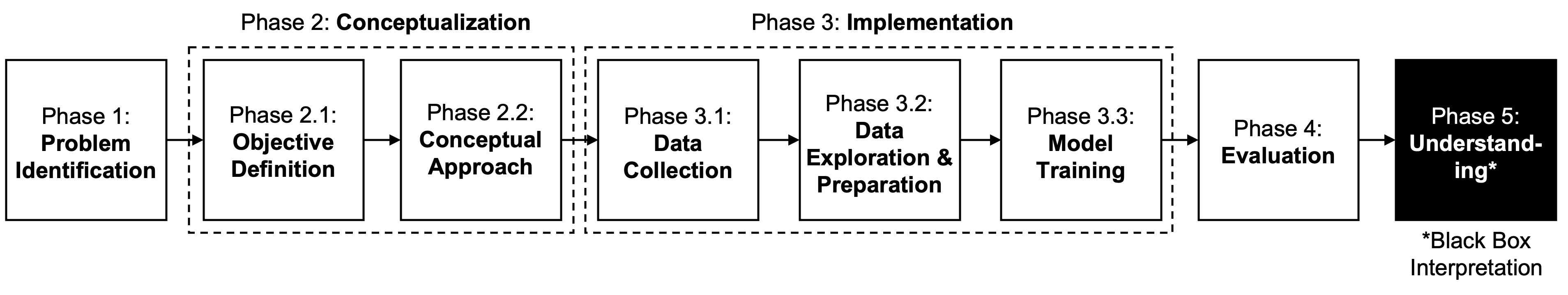}
  \caption{Steps in developing and understanding prediction-oriented black box ML artifacts}
  \label{fig:fig5}
\end{figure}

To create a methodology that guides DSR efforts in building and interpreting prediction-oriented ML artifacts, we unify the strengths of the methodologies from DSR and predictive modeling. In doing so, we build upon the phases of a DSR project as derived from the DSR methodologies. As shown in Figure \ref{fig:fig3}, these phases are also applicable to the methodologies from predictive modeling. Yet, the DSR methodologies put more emphasis on the execution of the conceptualization phase, whereas the methodologies from predictive modeling focus more on the implementation phase. We not only consider this in our methodology, but also put strong emphasis on black box model interpretation by providing guidance on how to understand a black box model using SHAP. Figure \ref{fig:fig5} depicts the steps of our proposed methodology.

\subsection{Phase 1: Problem Identification}

In the first phase of a DSR project that aims to develop an interpretable prediction-oriented ML artifact we follow the recommendations of all the DSR methodologies as well as BDAG by proposing to motivate and define a research problem. As shown in Figure \ref{fig:fig4}, such problem typically arises from the environment and forms the basis of a DSR project. However, theoretical knowledge might also serve to substantiate and highlight the relevance of a problem. Problem definition is crucial for project execution, as it guides the development of an effective artifact by providing valuable grounding for conceptualization and evaluation. Moreover, this phase has a strong justificatory purpose and serves to motivate the pursuit of a solution towards an audience \cite{Peffers2007}. After the problem definition, researchers may also formulate one or more research questions that may be answered throughout artifact evaluation and/or interpretation. The formulation of such research question(s) might explicitly take into account that theoretical contributions may arise inductively from data and not compellingly from mere deductive reasoning \cite{Muller2016}.

\subsection{Phase 2: Conceptualization}

In order to facilitate conceptualization, we divide this phase into two steps which we synthesize from the DSR methodologies and SBEM, namely (i) \textit{objective definition} and (ii) \textit{designing a conceptual approach}. Although these two steps do not equally apply to all of the DSR methodologies, they are crucial within the methodologies in which they appear, as they account for the definition of (i) the solution space and (ii) the solution itself (i.e., the artifact), respectively. For this reason, we constitute that dividing conceptualization into these two steps highly supports developing, structuring, and presenting a concept.

\subsubsection{Phase 2.1: Objective Definition}
In the first step, objectives have to be inferred from the problem definition with respect to the feasibility of a possible solution \cite{Peffers2007}. As for interpretable prediction-oriented ML artifacts, objective definition should commonly take two aspects into account: (i) one that applies to the practical utility of the prediction itself with respect to the given problem and (ii) one that pertains to the interpretation of a black box model and, thus, to the understanding of the phenomenon at hand. To this end, the artifact needs careful specification of what needs to be predicted, as this accounts for the measurement of the phenomenon of interest and impacts the choice of methods to be used throughout implementation \cite{Shmueli2011}. This specification includes determining whether the artifact seeks to solve a classification or regression problem, that is, the specification as to whether the target variable at hand is categorical or numerical. As for categorical outcomes, an alternative approach is to rank new observations according to their belonging to a certain class, which is referred to as \textit{ranking} \cite{Shmueli2011}.

\subsubsection{Phase 2.2: Conceptual Approach}

In the second step of conceptualization, the defined objectives serve to build a conceptual approach in the sense of a solution’s system architecture, as proposed by Nunamaker et al. (1990) \cite{Nunamaker1990}. Such an approach may provide a summary of the functionalities of the artifacts by putting its components, their relationships, and their interactions into perspective. Apart from system components, the conceptual approach may also include the interaction between the artifact and other entities, such as data sources, human stakeholders, or other artifacts. Thus, by clearly defining how the artifact aims to realize the objectives in the light of the given problem, such a conceptual approach particularly serves as a road map for implementation and evaluation. Moreover, the conceptual approach is well-suited to highlight the innovation of a solution. For this purpose, we recommend to build upon the \textit{DSR knowledge contribution framework} put forward by Gregor and Hevner (2013) \cite{Gregor2013}. The framework divides knowledge contributions (i.e., innovation) essentially into three types: (i) \textit{improvements} (i.e., new solutions for known problems), (ii) \textit{exaptation} (i.e., extension of known solutions to new problems), and (iii) \textit{inventions} (i.e., new solutions for new problems). It is illustrative to visualize the conceptual approach by using a corresponding figure.

\subsection{Phase 3: Implementation}

In line with all the predictive modeling methodologies, we divide implementation into three sub-phases: (i) \textit{data collection}, (ii) \textit{data exploration \& preparation}, and (iii) \textit{model training}.

\subsubsection{Phase 3.1: Data Collection}

\textbf{Data Sources.} Although it is desirable in the course of a DSR project, which aims to develop a data-driven artifact for computing empirical predictions, to draw upon rich and relevant real-world data, many research scenarios do not have access to such data and therefore depend on the use of alternative data sources. Beyond that, even in cases where data can be gathered from practice, predictive modeling often benefits from enriching the available data with data from complementary sources. For example, Hanke, Hauser, Dürr, \& Thiesse (2018) \cite{Hanke2018} develop a stationary product recommendation system for smart fashion retail environments using practitioner data (i.e., customer purchase history) enriched with publicly available contextual information (i.e., weather data from weather stations close to the specific stores under examination). Their results show that the incorporation of such open data leads to considerable improvements in the predictive accuracy of their models. Another project in the context of stationary fashion retailing, which even entirely refrains from using real-world data to develop a data-driven artifact, is presented by Hauser, Günther, Flath, \& Thiesse (2019) \cite{Hauser2019}. The authors develop an RFID-based automated checkout system which is capable of predicting product purchases of individual customers within fashion stores. Their artifact builds upon data (i.e., walking paths) gathered from a laboratory experiment, which was designed to mimic real world scenarios. Their results provide valuable insights into the opportunities and challenges that may come with the implementation of such system in the real world. The studies of Hanke et al. (2018) \cite{Hanke2018} and Hauser et al. (2019) \cite{Hauser2019} are only two striking examples to highlight that data-driven DSR projects may consider the use of various data sources, depending on availability and suitability. However, we argue that particularly suited data sources to drive a prediction-oriented ML artifact may be, but are not limited to:
\begin{itemize}
    \item Practitioner data (e.g., sales data, financial data, business process data, sensor data)
    \item Publicly available data sets (e.g., from Google dataset search)
    \item Publicly available data from APIs or websites (e.g., gathered via web scraping)
    \item Data from laboratory experiments
    \item Data from crowdsourcing (e.g., Amazon Mturk or \url{www.prolific.co})
    \item Simulated data (e.g., from Monte Carlo simulations)
\end{itemize}

\textbf{Sample Size.} Another important aspect of data collection is sample size. Most authors agree that sample sizes should be larger in predictive modeling compared to explanatory modeling \cite{Agarwal2014,Muller2016,Shmueli2011}. Shmueli and Koppius (2011) \cite{Shmueli2011} point out that this is due to four major reasons: (i) as opposed to estimating population-level parameters in explanatory modeling, predicting individual observations has a higher degree of uncertainty and therefore requires more observations; (ii) data-driven algorithms seek to build the structure of an empirical model by learning from data—the more data there is, the more there is to be learned; (iii) as predictive modeling often aims to capture complex relationships, larger samples may help to reduce model bias and sampling variance; (iv) more data are required for the creation of a test set that serves to evaluate the predictive accuracy of a model. However, in some situations—depending on the nature of the data, the problem at hand, and the methods used—even smaller samples may serve to train a model that possesses a sufficient degree of predictive power. Dürr, Griebel, Welsch, \& Thiesse (2020) \cite{Durr2020}, for example, conducted a study that utilizes texts from whitepapers to predict whether corresponding initial coin offerings are fraudulent or not. Although their final training sample—due to their under-sampling strategy—only consists of 166 observations, the authors are able to train an ML model that predicts fraud with a precision of 80\% on their test set. Thus, we conclude that, although larger sample sizes are preferable when it comes to training a prediction-oriented ML artifact, the utility of such artifact is determined by the accuracy of its out-of-sample predictions and not by the size of the training sample. Hence, smaller samples may be used to train predictive ML artifacts whenever effective.

\textbf{Dimensionality.} As opposed to the widely held belief that high-dimensional data sets are a curse in statistical modeling, Breiman (2001) \cite{Breiman2001a}—in his prominent \textit{Statistical Science} article—points out that “dimensionality can be a blessing” in predictive analytics. He substantiates this statement by arguing that each additional feature may comprise additional information that might be valuable for prediction—or, in his words: “The more predictor variables, the more information.” Moreover, in line with Müller et al. (2016) \cite{Muller2016} and Shmueli and Koppius (2011) \cite{Shmueli2011}, we recognize high data dimensionality to be characteristic for many predictive analytics studies that employ advanced ML algorithms, as such algorithms are designed to find patterns within high-dimensional data. In contrast to explanatory modeling, Shmueli and Koppius (2011) \cite{Shmueli2011} emphasize that in predictive modeling not every feature within a data set needs to be justified by theory. We argue that such insistence on theory would even lower the chance of uncovering novel patterns. However, as opposed to research efforts that have no ambition on gaining insights from predictive black box ML modeling—e.g., because their focus is on producing particularly accurate predictions, even at the expense of model interpretability—we recommend to be careful when it comes to incorporating features whose meaning is unclear. Such features are prone to providing opaque explanations to a given prediction.

\textbf{Data Types.} The proliferation of deep learning led to algorithms, such as convolutional neural networks, that are capable of processing various data types. Thus, prediction-oriented ML artifacts are not restricted to build upon traditional tabular data. Other possible data types may be texts, images, audio spectrograms, or videos \cite{LeCun2015}. However, although modern approaches to explainable AI support the interpretation of models that use other than tabular data, we recommend to verify the level of interpretability of the data under consideration before developing the artifact.

\textbf{Synthetic Data.} According to Gartner (2020) \cite{GartnerInc.2020}, “[s]ynthetic data is generated by applying a sampling technique to real-world data or by creating simulation scenarios where models and processes interact to create completely new data not directly taken from the real world.” Thus, algorithms that generate synthetic data are a promising means to (i) increase sample size or (ii) simulate whole data sets in cases where data is sparse. By now, these algorithms are already capable of generating various types of data, such as images or tabular data. Besides more traditional approaches to creating synthetic data, such as Monte Carlo simulations, modern examples of such algorithms are data augmentation techniques or generative adversarial networks \cite{Brock2018,Goodfellow2014,Xu2019,Xu2018}.

\subsubsection{Phase 3.2: Data Exploration \& Preparation}

This phase aims to (i) gain a comprehensive understanding about the data at hand and (ii) preprocess the raw data into convenient data sets. As opposed to SBEM, where data preparation comes before data exploration and variable selection, we argue that these three steps are difficult to separate from each other since they are heavily interwoven. Typically, this phase comprises an iterative procedure of exhaustively browsing, visualizing, understanding, selecting, and altering data. Therefore, in practice, this step is usually particularly error-prone and often time-consuming \cite{Wirth2000}. Moreover, this phase, in some cases, may result in more than one plausible final feature set. In such cases, the predictive performances of models trained using alternative feature sets may be compared in the evaluation phase.

\textbf{Data Partitioning.} According to Shmueli and Koppius (2011) \cite{Shmueli2011}, the available data need to be partitioned into three parts: (i) a training set, (ii) a validation set, and, finally, (iii) a test set. While the first serves to fit ML models, the second is used to facilitate algorithmic parameter adjustment (i.e., \textit{hyper-parameter tuning}) and, thus, model selection. The test set, however, is only used in the evaluation phase to assess model generalizability, that is, predictive out-of-sample performance of the selected model(s). Alternatively, instead of using a validation set, cross-validation may be employed during model training to enable hyper-parameter tuning and model selection. Even in such situations, it remains necessary to create a test set for evaluation. Typically, the test set comprises around 5-30\% the size of all the available data.

\textbf{Visualization.} According to Fayyad, Grinstein, \& Wirse (2002) \cite{Fayyad2002}, descriptive data visualization can support analysts in finding patterns that facilitate further analyses. In their words (p. 21):

\setlength{\leftskip}{1cm} \setlength{\rightskip}{1cm} \textit{A visualization can provide a qualitative overview of large and complex data sets, can summarize data, and can assist in identifying regions of interest and appropriate parameters for more focused quantitative analysis. In an ideal system, visualization harnesses the perceptual capabilities of the human visual system.}

\setlength{\leftskip}{0pt} \setlength{\rightskip}{0pt}

However, as data visualization comprises a vast number of techniques and depends highly on the nature of the underlying data, neither scholars nor practitioners have yet developed a standard process to data visualization (and probably never will). Thus, according to Shmueli and Koppius (2011) \cite{Shmueli2011}, exploratory data visualization should rather be understood as a “free-form fashion” to pattern recognition. Hence, the quality, suitability, and, ultimately, utility of data visualization heavily depend on the analytical and creative skills of an analyst. Exemplary methods that support data visualization are line graphs, scatter plots, isosurfaces, rubber sheets, volume visualization, and scalar glyphs \cite{Fayyad2002}.

\textbf{Data Cleansing.} Unclean data sets can be a serious issue when it comes to building a predictive ML artifact, as these may lead to (i) a biased model (in case the training set is unclean), (ii) a biased evaluation (in case the test set is unclean), or (iii) even both. Here, unclean data may consist of missing values, duplicate entries, measurement noise, typographical errors, or systematic errors \cite{Witten2011}. Before an analyst can cope with erroneous data, the errors should be unveiled. For this purpose, data visualization may serve as a means to detect anomalies within the data at hand. As opposed to duplicate entries, typographical errors, and systematic errors—which are typically relatively easy to fix once detected, i.e., through performing basic operations, such as deleting or adjusting respective entries—, missing values and measurement noise are sometimes challenging to handle. As for missing data, it should first be clarified if the data is informative of the to-be-predicted target \cite{Ding2010}. Thereupon, methods such as removing observations, removing variables, using proxy variables, creating dummy variables that indicate missing entries, and algorithms for imputation may be applied to handle missing values \cite{Shmueli2011}. With regard to measurement noise—although it is desirable in most scenarios to have as little noise as possible—, Witten et al. (2011) \cite{Witten2011} point out that it may be helpful to add artificial noise to features within the training data in cases where the corresponding features within the test set are noisy; however, this does not apply to target noise—here, it is best to train on noise-free instances.

\textbf{Feature Engineering.} Feature engineering is the process of transforming raw data into formats that are suitable for ML model training \cite{Zheng2018}. It differs from data cleansing in that it aims to process clean data instead of cleaning erroneous data. Similar to data visualization, feature engineering comprises an enormous number of methods, such as text mining, normalization, standardization, one hot encoding, dimension reduction, and much more \cite{Aggarwal2012,Zheng2018}. Moreover, feature engineering may consider combining available data with contextual information (e.g., from open data sources) to build new features. For instance, Teubner, Hawlitschek, \& Dann (2017) \cite{Teubner2017}, explain prices for homes on Airbnb by incorporating the independent variable \textit{distance to city center}, which they calculate using (i) the homes geo location coordinates from the Airbnb website and (ii) the respective coordinates from the corresponding city centers. As a result, the authors unveil the feature \textit{distance to city center} to have a significant impact on Airbnb prices. Thus, the utility of feature engineering—again—, similar to data visualization, highly depends on the analytical and creative capabilities of an analyst. Apart from that, domain and theoretical knowledge may also serve to facilitate feature engineering. As opposed to predictive modeling that refrains from model interpretation, we recommend building interpretable features that can be explained to an audience. As for dimension reduction, for example, this means that the utilization of techniques which produce rather uninterpretable features, such as Principal Component Analysis (PCA), should be avoided. Following the guidelines of the oft-cited article by Costello \& Osborne (2005) \cite{Costello2005}, we recommend using other forms of Exploratory Factor Analysis (EFA) instead of PCA in cases where features are to be compressed into fewer features. This means, we particularly recommend the use of EFA methods that employ oblique rotation over traditional PCA that utilizes orthogonal rotation, even though the former are prone to result in correlating features. However, such features from EFA have the advantages of (i) being more interpretable and, often, (ii) better capturing the initial features, particularly in cases where features reflect human behavior, as, in such cases, “we generally expect some correlation among [features], since behavior is rarely partitioned into neatly packaged units that function independently of one another” \cite{Costello2005}. Beyond that, advanced ML algorithms are capable of handling multi-collinearity. Subsequently, in order to calculate factor scores (i.e., feature values) based on EFA results, we recommend to follow the guidelines of DiStefano, Zhu, \& Mîndrilǎ (2008) \cite{DiStefano2008}. The authors describe several methods that serve the purpose of computing factors scores, i.e., (i) several types of sum scores, (ii) regression scores, (iii) Bartlett scores, and (iv) Anderson-Rubin scores. Although EFA-based factor scores are usually used in explanatory modeling and—in some situations—may lead to models that possess lower predictive accuracy compared to models trained on PCA scores, we point out that the development of interpretable prediction-oriented artifacts sits at the nexus of explanatory and predictive modeling. In such cases, “predictive analytics can focus on predictors […] that produce a relatively transparent model, while perhaps sacrificing some predictive power” \cite{Shmueli2011}.

\textbf{Feature Selection.} Feature selection (from a clean training data set) can roughly be divided into (i) theory-driven and (ii) data-driven approaches. Whereas the former focus on applying theoretical and domain knowledge to select features (as is common in explanatory modeling), the latter seek to select features by performing data analyses. Many modern ML algorithms do not require such a priori feature selection, as they are designed to find important predictors within a set of features autonomously. However, in some situations, model training can benefit from feature selection, e.g., in cases where the goal is to develop a particularly parsimonious model or if limited computational resources lead to slow training. Data-driven approaches to facilitate feature selection include correlation analyses \cite{Shmueli2011}, lasso regression (i.e., L1 regularization) \cite{Nagpal2017,Ng2004}, and tests that employ feature selection metrics, such as chi-square, information gain, or odds ratios \cite{Zheng2004}.

\subsubsection{Phase 3.3: Model Training}

In the last step of implementation, one or more ML models are fitted to the training data. To this end, appropriate ML algorithms need to be selected. As for regression or classification scenarios that employ tabular data a wide range of popular black box algorithms exist. Among these are support vector machines \cite{Vapnik1995}, deep neural networks \cite{Howard2020}, and ensemble methods, such as random forests \cite{Breiman2001}, or boosted trees \cite{Chen2016}. Thus, it is common in predictive modeling based on tabular data to train various alternative algorithms on a given training set to compare their predictive performances in the evaluation step. Here, aside from black box ML algorithms, simpler white box algorithms, such as linear or logistic regression may be trained to serve as predictive benchmarks. With regard to other than tabular data, the choice of algorithms is typically more limited to deep learning techniques \cite{Howard2020,LeCun2015} and may often result in only one or few suitable algorithms to address a given problem.

However, selecting a suitable algorithm does not automatically result in a well-performing model, even if a convenient training set is available. That is because even state-of-the-art ML algorithms—although usually less prone to over-fitting compared to more traditional statistical methods—bear the risk of over-fitting their respective training data. Over-fitting occurs whenever a model exhibits a particularly good fit with the training data, which, on the downside, comes at the expense of its generalizability, i.e., the accuracy of its out-of-sample predictions. To assess over-fitting, the predictive performance on the training set may be compared to the performance on the validation set. If the model lacks performance on the latter, the given algorithm’s hyper-parameters may be adjusted to improve out-of-sample predictions, if possible. For this purpose, strategies such as grid search or random search may be employed \cite{Bergstra2012}. Alternatively, in projects that refrain from creating and using a validation set, hyper-parameter-search may be combined with k-fold cross-validation to verify the predictive accuracy of a model:

\setlength{\leftskip}{1cm} \setlength{\rightskip}{1cm} \textit{K-fold cross-validation is an out-of-sample approach: a sample is split into k parts, where k-1 parts are used as the training sample (for model estimation) and the k-th sample is designated out-of-sample. This process is repeated k times, each time designating a different subsample as the out-of-sample part \cite{Shmueli2016}.}

\setlength{\leftskip}{0pt} \setlength{\rightskip}{0pt}

For each algorithm under consideration, the process of adjusting parameters and validating predictive out-of-sample performance of the respective model is repeated until, ideally, a relatively generalizable model is found. Ultimately, this model is selected to be tested on the withheld test data set in the evaluation phase.

\subsection{Phase 4: Evaluation}

In order to provide guidance on how to conduct evaluations in DSR projects, Venable, Pries-Heje, \& Baskerville (2016) \cite{Venable2016} propose a \textit{framework for evaluation in design science} (FEDS). In their framework, the authors suggest specifying DSR evaluation efforts along two dimensions: (i) functional purpose of the evaluation and (ii) its paradigm. We now elaborate on these two dimensions and their transferability to our methodology.

\textbf{Functional Purpose.} According to FEDS, the functional purpose of an evaluation specifies \textit{why to evaluate} and may either be (i) \textit{formative} or (ii) \textit{summative}. According to Lee and Hubona (2009) \cite{Lee2009}, formative validity is “an attribute of the \textit{process} by which a theory [or an artifact] is […] built” and summative validity is “an attribute of the \textit{sum result} or \textit{product} of the process”. Accordingly, in FEDS, formative evaluations may help to improve the process of artifact development and, thus, the artifact itself, whereas summative evaluations seek to judge the extent to which an artifact matches expectations (i.e., the utility of the final artifact).

When it comes to building and evaluating prediction-oriented ML artifacts, however, the alternating process of hyper-parameter-tuning and validating the predictive performance of a model during training comprises multiple formative evaluation steps. This means that each assessment of the predictive performance of a model on a corresponding validation set (or in k-fold cross validation) is—according to FEDS—equal to a formative evaluation step, as it serves as “feedback” to improve the process of artifact development (i.e., model training). Thus, as such formative evaluation is already considered in phase 3.3 of our methodology, and in line with all of the methodologies presented in section \ref{sec:2}, we propose to perform a summative evaluation in the actual evaluation phase, i.e., one evaluation per considered model using the test set.

\textbf{Paradigm.} Aside from the functional purpose of an evaluation, FEDS distinguishes between two paradigms: (i) \textit{artificial} and (ii) \textit{naturalistic} evaluation. The chosen paradigm specifies \textit{how to evaluate}. In essence, artificial evaluation refers to evaluation efforts that are conducted beyond the real environment, such as laboratory experiments, simulations, and mathematical proofs. In contrast, naturalistic evaluation aims to capture the performance of an artifact in its real environment (e.g., an organization) and typically utilizes methods such as case studies, field experiments, and surveys.

Consequently, as for prediction-oriented ML artifacts, the extent to which an evaluation is rather artificial or naturalistic depends on the source of the holdout data (i.e., the validation set for formative evaluations and the test set for summative evaluations). In case the holdout set is simulated or comes from a laboratory experiment, the corresponding evaluation is more artificial; in contrast, if the holdout set comes from the real environment, the corresponding evaluation is more naturalistic. In both cases, the evaluation phase aims to assess model generalizability, that is, the accuracy of out-of-sample predictions using suitable metrics, such as those mentioned in section \ref{sec:2.2}. Moreover, in this phase, the predictive performance of at least one ML model may be compared with competing models or at least one suitable benchmark.

\begin{figure}[H]
  \centering
  \includegraphics[width=1.0\textwidth]{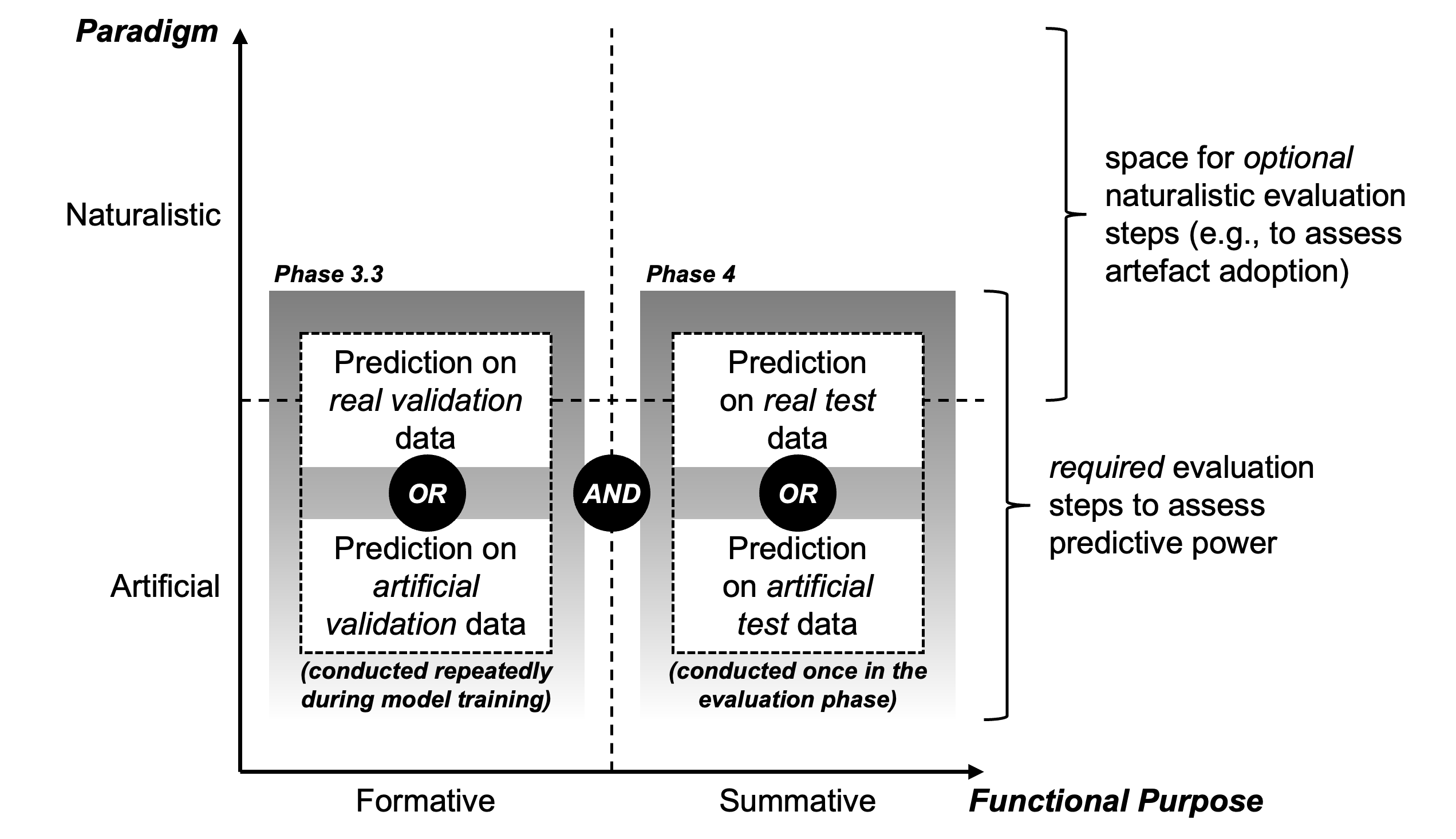}
  \caption{FEDS for prediction-oriented artifacts}
  \label{fig:fig6}
\end{figure}

However, we contend that all evaluations that solely strive to assess the predictive accuracy of an ML model are rather artificial. We justify this view by arguing that, in the real world, more aspects are important for evaluation than mere predictive power, such as the actual—or at least intended—use of an artifact by relevant practitioners. That means that, even in cases where an artifact may possess substantial predictive power, it cannot exploit its utility if it fails adoption in the real world. By now, technology adoption research has repeatedly shown that the use of an artifact is not compellingly caused by its objective usefulness, but rather its perceived usefulness, its perceived ease of use, social influence, habit, trust, and other antecedents \cite{Davis1989,Gefen2003,Venkatesh2012}. Thus, a comprehensive naturalistic evaluation would consider (i) more aspects that are aimed at investigating adoption and (ii) respective human behavior and/or opinions. However, in this paper, we refrain from exhaustively examining naturalistic evaluations, as these are neither required for proving the predictions of an ML artifact accurate nor to interpret a black box ML model. Still, studies that aim to develop predictive ML artifacts may consider conducting optional naturalistic evaluations—particularly once artificial evaluations and model interpretations can be presented to an audience (i.e., after performing all the steps of our methodology), as these most likely affect artifact adoption. Here, we especially expect model interpretability to foster adoption. Figure \ref{fig:fig6} summarizes our elaborations on ML artifact evaluation.

\subsection{Phase 5: Understanding}
\label{sec:4.5}

After evaluation, at least one model needs to be selected for interpretation. We recommend opting for the model with the highest predictive accuracy that still builds upon interpretable features. Furthermore, we deliberately discourage analysts from interpreting models with low predictive power, as this will certainly result in unreliable explanations. Moreover, it is conceivable that, in this phase, interpretations from competing models are compared to verify consistency of feature importance across models. As outlined before, to understand a black box model, we utilize SHAP as an instance of explainable AI tools.

Once a model is chosen, explainable AI tools commonly allow for uncovering (i) local feature effects for single observations, (ii) global feature effects for an entire model, (iii) feature interaction, and (iv) data type specific patterns. For these purposes, SHAP offers respective visualizations, which we describe in the following according to their documentations\footnote{The visualization types are documented at \url{https://github.com/slundberg/shap} and \url{https://shap.readthedocs.io/en/latest/index.html}, which we accessed on the 23\textsuperscript{rd} of September, 2020.}. Typically, to calculate SHAP values, the training data set is used \cite{Lundberg2020}. However, in some exceptional cases, it may be reasonable to calculate SHAP values based on other data sets, such as the test set, e.g., if local predictions in the test data need to be explained. Yet, apart from such cases, we recommend drawing upon the training data when it comes to model interpretation.

\textbf{Force Plots.} Force plots are designed to explain local predictions. To this end, force plots for single observations—and their similar waterfall plots, to which we also refer as force plots—visualize the impact of each feature on the model output, that is, the extent to which each feature contributes to push the corresponding prediction from the base value (i.e., the average model output). In this context, every feature may either exert (i) a positive, (ii) a negative, or (iii) no impact on a corresponding prediction. Force plots further order features by their respective importance for a given prediction. Consequently, such plots are particularly suited to explain how and why individual predictions occur. In addition, SHAP allows for visualizing multiple stacked force plots in one compound plot. Thereby, explanations for an entire data set can be visualized. This approach is particularly suitable for exploring clusters of similar observations within a respective data set (i.e., observations with similar model inputs and outputs) \cite{Lundberg2017a}.

\textbf{Summary Plots.} As opposed to force plots, summary plots are rather suited to gain a global understanding of model-wide feature importance. For this purpose, SHAP supports the visualization of average SHAP value magnitudes per feature over an entire data set in the form of a summarized bar plot, which sorts features in descending order of global importance. Such type of summary plot is especially valuable in situations that require an easily comprehensible overview of important factors that are commonly associated with the phenomenon of interest. Aside from bar plots, SHAP offers another type of summary plot (i.e., bee swarm plots) that visualizes the SHAP values of each feature for every observation in the underlying data set. Although this plot also sorts features by their global importance, it provides deeper insights into the impact of each feature on the model output, that is, the distribution of low/negative and high/positive SHAP values across all observations. Therefore, this plot may particularly serve to investigate whether the relationship between a feature and a target is positive/negative or linear/non-linear.

\textbf{Dependence Plots.} These plots are a valuable means to (i) understand the effect of a single feature on the model output and (ii) to uncover interaction effects between features. Dependence plots achieve the former by plotting the actual values of a given features against their corresponding SHAP values in a scatter plot. Moreover, to investigate interactions, the points within the plot can be colored according to the values of a second feature. This allows for uncovering bivariate interactions, i.e., the extent to which the model output is sensitive to changes in the values of two interacting features.

\textbf{Decision Plots.} Decision plots are a powerful tool with which to support various types of analyses. They allow to visualize the path of decisions, which is undergone when the feature values of an observation pass through a given model when computing a respective prediction, starting from the base value. Decision plots support (i) visualizing the paths of single and multiple observations from one or more models and (ii) processing SHAP as well as interaction values. As a consequence, according to the documentation, such plots may serve to (i) show a large number of feature effects clearly; (ii) visualize multi-output predictions; (iii) display the cumulative effect of interactions; (iv) explore feature effects for a range of feature values; (v) identify outliers; (vi) identify typical predictions paths; (vii) and compare predictions for several methods.

\textbf{Heatmaps.} Heatmaps combine several aspects of force plots for entire models and summary plots. Besides showing feature importance across an entire data set, heatmaps visualize (i) model outputs and (ii) SHAP values for each feature and observation. Moreover, heatmaps use coloring to highlight feature effects according to their corresponding SHAP values. In doing so, SHAP orders observations in a way that visualizes aggregations of similar SHAP values as visible clusters. Apart from supporting the uncovering of clusters, such heatmaps serve to explain both local and global feature importance.

\textbf{Data Type Specific Plots.} These plots represent a special case of visualization supported by SHAP, as they are restricted to visualize SHAP values for specific data types. On the upside, these types of plots are tailored for uncovering data type specific patterns. Currently supported data types for data type specific plots are texts and images. Image plots, on the one hand, serve to assign SHAP values to the pixels of a given image and, ultimately, color these pixels according to these values. This allows to highlight regions on an image that are important for prediction. These plots are applicable in cases where image data serves to train deep learning models. On the other hand, text plots allow for highlighting words and text regions that are important in prediction-oriented text mining.

As mentioned in section \ref{sec:2.3.1}, we hereby—again—point out that associations between a feature and the output of a model do not inherently represent causal relationships, even in cases where a feature exerts a strong impact on the target. Still, the investigation of such association-based explanations is highly valuable for gaining insights into the nature of a given phenomenon and, thus, to inform both theory and practice, particularly because these associations—as being the result of predictive modeling—already proved to possess a certain degree of generalizability. Therefore, such associations should be understood as generalizable patterns that hold the potential to indicate causality.

\section{Illustrative Use Case}

After the development and description of a methodology that is suitable for guiding DSR efforts in building and interpreting predictive ML-driven artifacts, we now showcase its application. For this purpose, we consider an exemplary use case from the sharing economy. More precisely, we pass through each step of our methodology in order to build and interpret an artifact that computes price recommendations for Airbnb listings. These recommendations are intended to support property owners in finding a suitable rental price per night when first entering the market. Before artifact development, in line with our methodology, we start by motivating and defining the corresponding research problem and formulate proper research questions.

\subsection{Problem Identification}

Airbnb is a showpiece in the sharing economy comprising over 7 million listings in over 100,000 cities and more than 150 million users worldwide \cite{IPropertyManagement2020}. Consequently, Airbnb’s value exceeded \$113 billion in 2021 with the company’s yearly revenue being around \$5.992 billion \cite{MacrotrendsLLC2022}. Since the company’s business model builds upon providing a platform that brings together (i) property owners who want to rent out their homes for short time periods (\textit{hosts}) and (ii) guests, it is not only Airbnb that makes profit from their platform, but also the property owners themselves. This led to an ever-increasing number of hosts who made a business from participating in Airbnb (currently around 650,000 worldwide) \cite{IPropertyManagement2020}. But, despite the great opportunity to profit from renting, participating in Airbnb can be challenging for hosts, especially when it comes to finding the right rental price per night. This difficulty becomes particularly evident for hosts who are new to the market for two reasons: (i) hosts may be unfamiliar with suitable pricing ranges and (ii) their user profiles lack important price determinants that come with increasing membership duration, such as an average rating score, number of ratings, and whether they are \textit{superhosts}\footnote{Hosts are typically awarded with the superhost badge in case they meet a range of pre-determined criteria, such as accommodating at least 10 guests within a year, receiving at least a share 80\% five-star ratings, and more.} or not \cite{Teubner2017}. With reference to the literature on recommender systems, we refer to this initial lack of information as \textit{cold-start problem}. As a consequence of this problem, a considerable number of tutorials in corresponding blogs and forums emerged which support hosts in finding the right prices on Airbnb \cite{AirbnbCommunity2016,AirbnbInc.2020,GuestyInc.2018,RentingYourPlace2020}. Although these tutorials may have a certain value, most of them are intended for hosts with some experience, provide instructions that are based on excessive manual research, or give rather vague price recommendations. Moreover, from a scientific viewpoint, there has been some research on Airbnb pricing—however, we are not aware of studies that particularly address the cold-start problem we described. Thus, we argue that a tool which helps hosts on Airbnb to find a reasonable starting price when they first enter the market would be both useful and innovative. Therefore, in the present exemplary study, we seek to address this issue by developing an ML-driven artifact, which is able to predict appropriate starting prices per night and to provide corresponding explanations. For this purpose, proper research questions are:

\setlength{\leftskip}{1cm} 
\textbf{RQ1: What might be a suitable design of an ML artifact that assists hosts in finding starting prices on Airbnb?}

\textbf{RQ2: Which insights can be drawn from the interpretation of the black box ML model driving the artifact?}

\setlength{\leftskip}{0pt} 

\subsection{Conceptualization}

\subsubsection{Objective Definition}

As per our formulated research questions, this exemplary study comprises two distinct objectives: (i) building an artifact capable of predicting accurate prices for homes on Airbnb and (ii) gaining insights into factors driving Airbnb pricing from the interpretation of the artifact. In order to meet the first objective, we train suitable ML models using actual data from Airbnb. In doing so, we identify the model that can predict average prices per night most accurately. We calculate these using a selection of prices from a one-year period. Thus, as the considered target variable is numerical, we solve a regression problem. Moreover, as the artifact we develop is intended to support hosts that are new to the market, we refrain from utilizing features associated with membership duration, such as the number of ratings, since the profiles of new members lack these kinds of information. Besides the exclusion of such features, we do not exclude other features from modeling a priori. With regard to our second objective, we employ SHAP to get insights into the ML model that proves to possess the greatest predictive accuracy. These insights may assist (i) hosts in finding settings that can be adjusted by them in order to charge higher prices per night and (ii) researchers in broadening their knowledge about price finding in the sharing economy, particularly on Airbnb.

\subsubsection{Conceptual Approach}

\begin{figure}[H]
  \centering
  \includegraphics[width=1.0\textwidth]{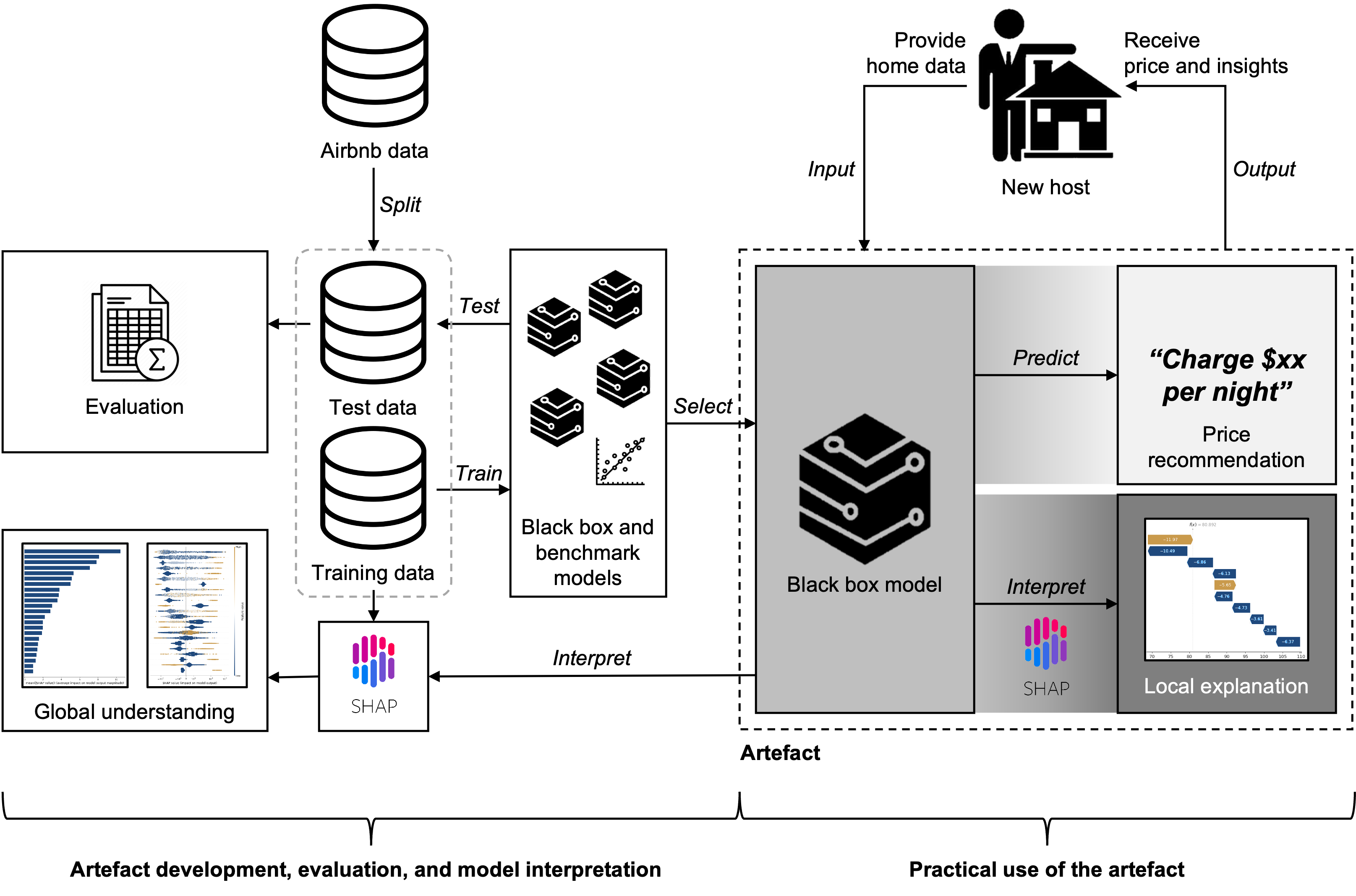}
  \caption{Conceptual approach}
  \label{fig:fig7}
\end{figure}

In order to meet the objectives, we propose the conceptual approach depicted in Figure \ref{fig:fig7}. The approach is divided into two parts: (i) the concept for artifact development, evaluation, and global model interpretation and (ii) the concept for the practical use of the artifact. The former illustrates how we make use of a data sample from Airbnb to develop the artifact. In essence, we split the available data into a test and a training set to build and evaluate multiple black box and benchmark models. Here, we opted for employing k-fold cross validation during model training instead of using a validation set. Once the models are trained, we select the model with the highest predictive accuracy for global interpretation. The insights gained from interpreting the entire model are particularly interesting for researchers who seek to understand price finding on Airbnb in general. Furthermore, the right part of the conceptual approach depicts how hosts may use the artifact once it is developed: A host may put the data of a home into the artifact to obtain a price recommendation and a local explanation. Ideally, the latter supports the host in making useful adjustments in the listing (e.g., upload more pictures) that contribute to justify an increase in price. As the proposed artifact is aimed at solving a known problem (i.e., the cold-start problem on Airbnb) in a new way (i.e., by calculating price recommendations and corresponding explanations automatically), it is an \textit{improvement} in the sense of Gregor and Hevner’s (2013) \cite{Gregor2013} DSR knowledge contribution framework.

\subsection{Implementation}

\subsubsection{Data Collection}
To collect Airbnb data, we used a suitable API, which is available on GitHub\footnote{\url{https://github.com/nderkach/airbnb-python}}. The API supports the collection of various data on homes in a given city, such as locations, amenities, home types, and more. We collected 50 homes from each of the 50 most populated cities in the USA, resulting in a total of 2,500 homes. Moreover, for each home, the API allows to collect prices per night for a period of one year in advance. We collected all available prices per night for each home. The data was gathered on April 30, 2020.

\subsubsection{Data Exploration \& Preparation}
\begin{figure}[H]
  \centering
  \includegraphics[width=1.0\textwidth]{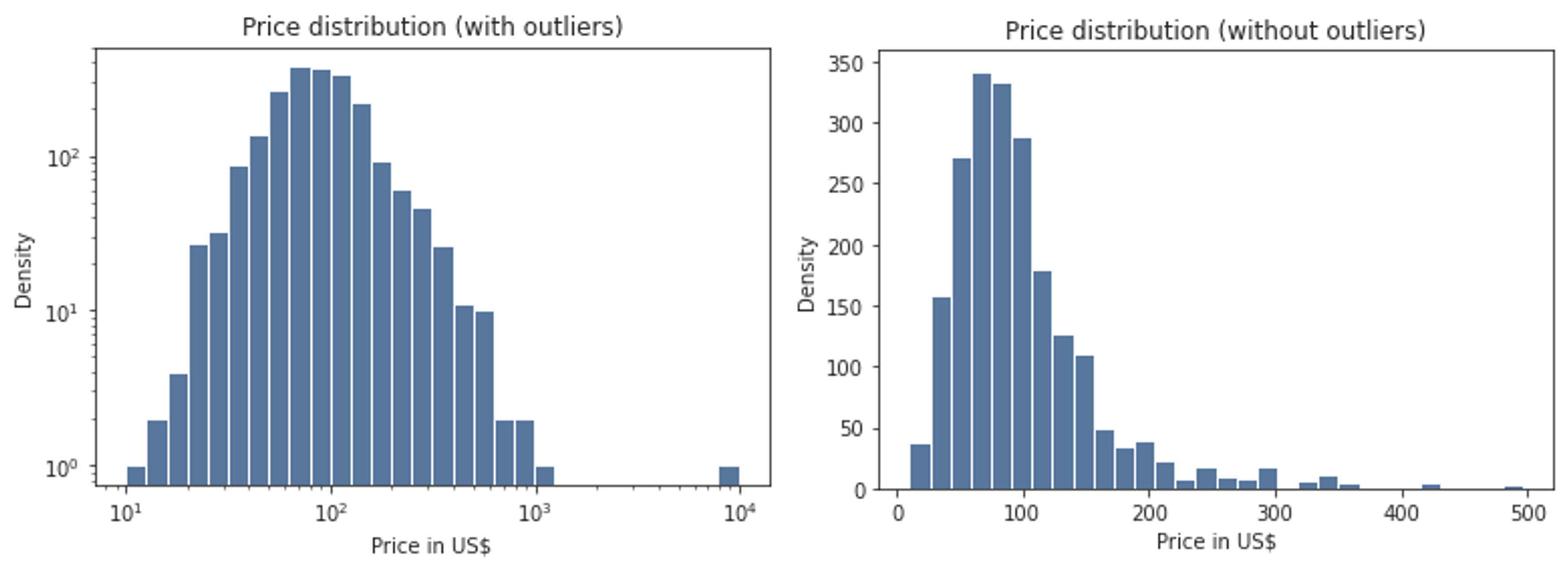}
  \caption{Price distribution}
  \label{fig:fig8}
\end{figure}

After data collection, we first reviewed the data for (i) missing values and (ii) duplicate entries\footnote{The data exhibited duplicate homes because we considered the boroughs of New York as distinct cities when collecting the data using the API. However, some homes were assigned to more than one borough on Airbnb, which is why some of these appeared more than once in our sample.}. In doing so, we deleted homes with at least one missing feature or target value to ensure data reliability. Moreover, we deleted duplicate homes. This procedure resulted in the exclusion of 388 homes. Subsequently, to get representative prices per night, we calculated average prices for each home. The left plot in Figure \ref{fig:fig8} depicts the corresponding price distribution. However, the prices comprised 16 homes (~0.76\%) with average prices over 500\$ per night, which we considered outliers. The right plot in the figure shows the price distribution without outliers—note that both distribution plots are illustrated using two different scales on both axes. Our further analysis builds upon the outlier-free data, which encompasses 2096 homes.

In a next step, we randomly split the available data into a training and a test set. While the former consists of approximately 90\% of the available data (i.e., 1886 homes), the latter comprises the remaining 10\% (i.e., 210 homes). As is common in predictive analytics, we conducted further analyses based on the training set. Thus, the test set solely serves to assess the predictive power of the trained models in the evaluation step.

After splitting the data, we performed various feature engineering steps, in particular, clustering amenities and home types; calculating distances to corresponding city centers; counting the number of foreign languages spoken by the host; computing the ratios (i) beds per bedroom, (ii) persons per bedroom, and (iii) persons per bathroom; and leveraging deep learning (i.e., convolutional neural networks) not only to classify preview pictures of each home as (i) a bathroom, (ii) a bedroom, (iii) a dining room, (iv) an exterior area, (v) an interior area, (vi) a kitchen, or (vii) a living room automatically \cite{Poursaeed2018}, but also to assign an MOS (mean opinion score) as a measure of picture quality to each of these preview pictures \cite{Hosu2020}. Moreover, we incorporated external data on the cities, where these homes are located, i.e., the population and population density. Subsequently, we utilized the L1 regularization which is implemented in extreme gradient boosting (XGBoost) \cite{Chen2016}, to select suitable features for model training. The final feature set consists of 47 features—we provide an overview of these in Appendix.

\subsubsection{Model Training}
For model training, we opted for using (i) two naïve dummy models that serve as predictive benchmarks, (ii) two white box methods, and (iii) three popular state-of-the-art black box algorithms. The first are strategies that use arithmetic mean and median values of the training prices to predict prices in the test set; the second are linear regression and a decision tree; the third are an artificial neural network as implemented in FastAI (with two layers), a random forest, and XGBoost. Moreover, we employed 5-fold cross validation combined with grid search to select hyper-parameters during the training of the white box and black box models.

\subsubsection{Evaluation}
After training, we utilized every model to predict prices in the test set. To evaluate the predictive power of each model, we calculated the mean average error (MAE), median absolute deviation (MAD), mean squared error (MSE), and the prominent R\textsuperscript{2} metric. The results are shown in Table \ref{tab:tab2}.

\begin{table}[H]
\centering
\caption{Evaluation results}
\label{tab:tab2}
\begin{tabular}{|lllll|}
\hline
\multicolumn{1}{|l|}{\textbf{Model}}                                                                                                         & \multicolumn{1}{l|}{\textbf{MAE}}                                                                                                           & \multicolumn{1}{l|}{\textbf{MAD}}                                                                                                          & \multicolumn{1}{l|}{\textbf{MSE}}                                                                                                            & \textbf{Out-of-sample R\textsuperscript{2}}                                                                                                        \\ \hline
\multicolumn{1}{|l|}{Dummy (mean)}                                                                                                           & \multicolumn{1}{l|}{39.77}                                                                                                                  & \multicolumn{1}{l|}{32.07}                                                                                                                 & \multicolumn{1}{l|}{2859.39}                                                                                                                 & -0.04                                                                                                                            \\ \hline
\multicolumn{1}{|l|}{Dummy (median)}                                                                                                         & \multicolumn{1}{l|}{37.83}                                                                                                                  & \multicolumn{1}{l|}{27.17}                                                                                                                 & \multicolumn{1}{l|}{2957.64}                                                                                                                 & -0.01                                                                                                                            \\ \hline
\multicolumn{1}{|l|}{LR}                                                                                                                     & \multicolumn{1}{l|}{31.54}                                                                                                                  & \multicolumn{1}{l|}{22.81}                                                                                                                 & \multicolumn{1}{l|}{1996.25}                                                                                                                 & 0.29                                                                                                                             \\ \hline
\multicolumn{1}{|l|}{DT}                                                                                                                     & \multicolumn{1}{l|}{31.99}                                                                                                                  & \multicolumn{1}{l|}{21.5}                                                                                                                  & \multicolumn{1}{l|}{2338.39}                                                                                                                 & 0.18                                                                                                                             \\ \hline
\multicolumn{1}{|l|}{ANN}                                                                                                                    & \multicolumn{1}{l|}{30.11}                                                                                                                  & \multicolumn{1}{l|}{21.07}                                                                                                                 & \multicolumn{1}{l|}{2085.01}                                                                                                                 & 0.27                                                                                                                             \\ \hline
\multicolumn{1}{|l|}{RF}                                                                                                                     & \multicolumn{1}{l|}{28.98}                                                                                                                  & \multicolumn{1}{l|}{20.81}                                                                                                                 & \multicolumn{1}{l|}{1819.85}                                                                                                                 & 0.36                                                                                                                             \\ \hline
\multicolumn{1}{|l|}{XGB}                                                                                                                    & \multicolumn{1}{l|}{\textbf{27.35}}                                                                                                         & \multicolumn{1}{l|}{\textbf{18.91}}                                                                                                        & \multicolumn{1}{l|}{\textbf{1741.47}}                                                                                                        & \textbf{0.39}                                                                                                                    \\ \hline
\multicolumn{5}{|l|}{\begin{tabular}[c]{@{}l@{}}\textbf{Models:}\\ Dummy (mean) predicts the test data using the target mean from the training data.\\ Dummy (median) predicts the test data using the target median from the training data.\\ LR := Linear regression\\ DT := Decision tree\\ ANN := Artificial neural network\\ RF := Random forest\\ XGB := Extreme gradient boosting\\  \\ \textbf{Metrics:} \\ MAE := Mean absolute error (lower values=better)\\ MAD := Median absolute deviation (lower values=better)\\ MSE := Mean squared error (lower values=better)\\ Out-of-sample $R^{2}$ := Coefficient of determination calculated using predictions/observations \\ from the test set (higher values=better) \end{tabular}} \\ \hline
\end{tabular}
\end{table}

The evaluation unveils the two dummy models to be the two worst choices to predict prices on Airbnb. However, due to their simplicity, this finding is rather unsurprising. More interesting is the considerable predictive power of the linear regression model which not only outperforms the dummy benchmarks, but also surpasses the decision tree in all metrics except for the MAD and even slightly beats the artificial neural networks MSE and R\textsuperscript{2}. Here, we particularly assume the lack in predictive accuracy of the neural networks to result from the size of the training set and argue that the algorithm would produce a superior model in a “big data” setting. However, both tree ensemble approaches provide predictions superior to those of all other models, with the XGBoost model outperforming the random forest, thus being the most accurate model.

As for the first of the research questions we formulated in the problem identification step, we can conclude that our conceptual approach, combined with the features we used, our employed training strategy, and the XGBoost algorithm constitutes a fairly suitable design for an artifact to predict starting prices on Airbnb. However, when recreating our artifact, developers might consider using a larger training data set, which could take listings from other countries or regions into account, depending on the corresponding field of application. Moreover, the creation of additional or refined features is conceivable. By using a larger training sample or more features, we would generally expect the XGBoost model to increase in predictive accuracy.

To answer the second of our research questions, we interpret the XGBoost model in the next step.

\subsubsection{Understanding}

As per the conceptual approach we put forward, we are interested in (i) gaining a global understanding of the XGBoost model and (ii) explanations of individual predictions that may serve to inform hosts.

\begin{figure}[H]
  \centering
  \includegraphics[width=1.0\textwidth]{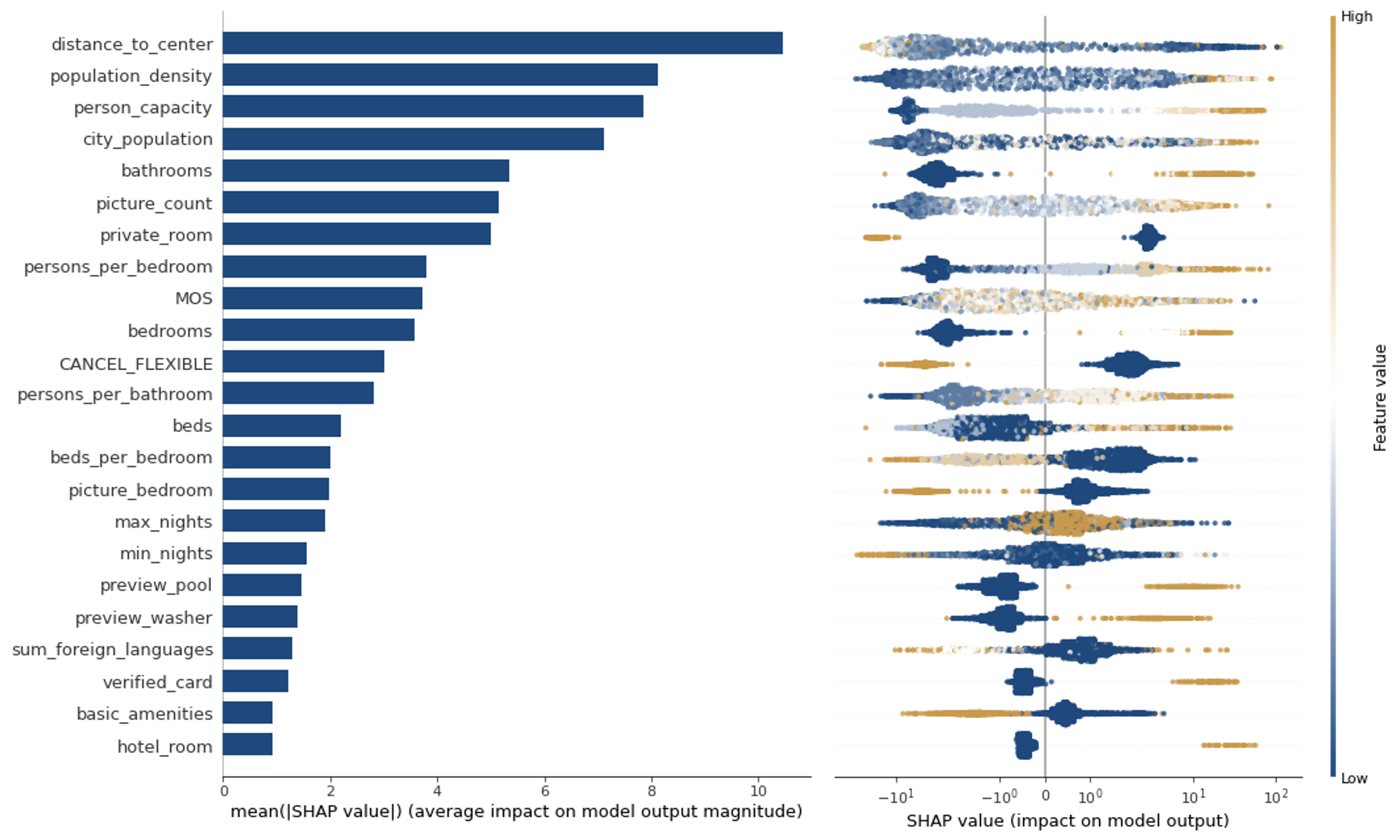}
  \caption{Global understanding}
  \label{fig:fig9}
\end{figure}

We hereby start with global understanding. For this purpose, we employ SHAPs summary plots, as described in section \ref{sec:4.5}. The left plot in Figure \ref{fig:fig9} depicts a bar summary plot, whereas the plot on the right side of the figure is a bee swarm plot. Both plots show the 23 most important features for prediction in descending order. The remaining features are comparatively negligible for price prediction. While the bars in the bar plot illustrate the average SHAP value magnitude of each feature, the bee swarm plot additionally shows the distribution of individual feature values across their corresponding SHAP values.

Figure \ref{fig:fig9} unveils the distance of a home to its corresponding city center (“distance\_to\_center”) as the most important feature to predict prices per night on Airbnb. Here, despite very few exceptions, medium and high distances exert a negative impact on the model output, whereas low distances tend to predict higher prices. The next most important features are the population density of a city (“population\_density”), a home’s capacity in terms of people that can stay overnight (“person\_capacity”), and a city’s total population (“city\_population”). In essence, all of these features tend to push the XGB models output towards lower prices if feature values are low and vice versa. Therefore, until here, we can conclude that large homes, which are close to the centers of big cities, are typically more expensive than other homes.

The remaining features may be examined in a similar fashion to gain a more complete understanding of the model. However, instead of investigating every effect, we focus on the most interesting ones:

\begin{itemize}
\item Number of uploaded pictures (“picture\_count”): An increasing number of pictures is associated with higher prices. This effect is not only the sixth most important feature that predicts the target in the XGBoost model, but it is also especially interesting as it is not determined by the inherent characteristics of a home or its location. Although we cannot reliably confirm a direct causal effect between the number of pictures and the price (i.e., both variables may also be the result of other causes), we argue that, in practice, it might help to upload more pictures to justify an increase in price—even though we cannot confirm causality, we cannot preclude it either.
\item Number of persons per bedroom/bathroom (“persons\_per\_bedroom”/”persons\_per\_bathroom”): As much as it would be conceivable that a lower number of persons per bedroom or bathroom was associated with higher prices (due to more space per person), the opposite (i.e., more persons per room leading to higher prices) would likewise be reasonable (due to a greater person capacity). According to the bee swarm plot, the latter is the case. Yet, these effects are heavily influenced by the corresponding home type (i.e., whether the home is an entire place or a private room). We provide an overview of these and other feature interactions in Appendix.
\item Quality of the preview picture (“MOS”): The quality of the preview picture, which we measured using the MOS as proposed by Hosu et al. (2020) \cite{Hosu2020}, is also considerably predictive of our target. Whereas lower quality pictures tend to predict lower prices, higher MOSs broadly push the model output towards higher prices. Here, we draw the same conclusions as for the number of pictures—although we cannot confirm a causal effect, we recommend hosts to use high quality preview pictures if they want to raise prices.
\item Flexible canceling (“CANCEL\_FLEXIBLE”): This effect is particularly interesting as the presence of a flexible canceling policy clearly exerts a negative influence on the model output, whereas other, stricter canceling policies contribute to increase prices. We argue that this most likely is not a direct causal effect, as it is meaningless—we would decisively expect flexible canceling to be perceived as convenient by guests. We rather find it conceivable that other factors prompt hosts to select flexible canceling for their low-price homes (e.g., to compensate for other disadvantages). Yet, this effect may have the potential to constitute a worthwhile avenue for future research.
\item The preview picture shows a bedroom ("picture\_bedroom"): Interestingly, having a bedroom as a preview picture exerts a negative impact on the price while other preview pictures (e.g., living room, kitchen, or exterior) have a positive or no impact on the target. Although this effect might be sample-specific, with our current state of knowledge we would recommend hosts to refrain from using bedroom pictures as preview pictures. Yet, this effect might be investigated further in the future.
\item Maximum/minimum number of overnight stays (“max\_nights”/”min\_nights”): While a high number of minimum nights required to rent a home are distinctly associated with lower prices, low values can either have a negative, positive, or no impact on the model output. Moreover, some medium values even push the model output towards high prices. Even more ambiguous is the effect of maximum nights—while high values tend to have less impact on the model output than low values, the latter either have (i) a negative or (ii) a strong positive influence on prices. Without further analyses, this effect is hardly explainable.
\item Number of foreign languages (“sum\_foreign\_languages”): Along with “max\_nights”, interpreting this effect would also require further analyses. It seems like a medium number of foreign languages have a negative impact on the target variable, while high values either have a strong positive or negative influence. Low values have no or a weak positive influence on prices.
\end{itemize}

While some of the global interpretations are highly informative (e.g., that more pictures are associated with higher prices), some are rather vague (e.g., that flexible canceling is associated with lower prices). However, besides the value of the former to serve as cornerstones for price finding on Airbnb, the latter may serve to draw researchers’ attention and, thus, prompt them to investigate these unclear effects in their future efforts. Moreover, our findings are well-suited to be incorporated in explanatory models that serve to explain prices on Airbnb or even the sharing economy as a whole.

\begin{figure}[H]
  \centering
  \includegraphics[width=1.0\textwidth]{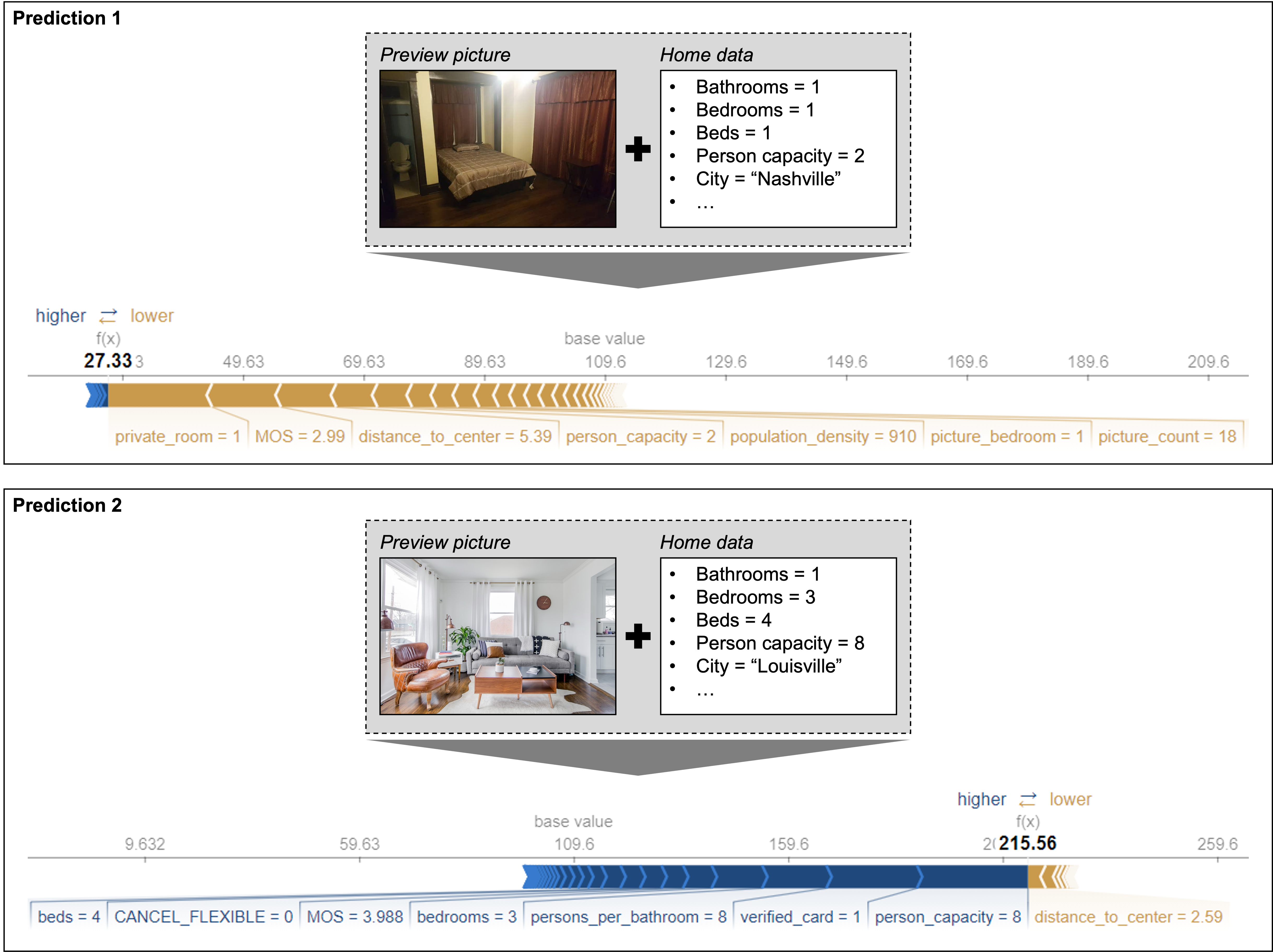}
  \caption{Local explanations}
  \label{fig:fig10}
\end{figure}

In addition to a global understanding, we next show how local predictions and explanations may be used to support individual hosts. For this purpose, we use SHAP’s force plots and consider two exemplary homes from our data: one with a lower price (prediction 1) and one with a higher price (prediction 2). Both predictions—along with their corresponding force plots—are depicted in Figure \ref{fig:fig10}. For each prediction, the figure shows a preview of the input a host would need to provide (i.e., the preview picture of a home and additional data required to build our features) and the composition of the corresponding prediction as per the force plot. Each force plot illustrates the XGBoost model’s (i) base value (i.e., the price the model would predict if no feature values were available), (ii) important feature values that contribute to push the actual model output from the base value (orange contributions increase the price/green contributions decrease the price), and the actual model output (i.e., the prediction) shown in bold between the orange and green contributions. We note that the prices are consistently given in US\$ and that both predictions are fairly accurate—while prediction 1 (27.33\$) exhibits an absolute error of -8.63\$ (i.e., the actual average price is 18.70\$), prediction 2 (215.56\$) has an error of 7.86\$ (i.e., the actual average price is 223.42\$).

As opposed to prediction 1, the listing of prediction 2 broadly exploits the range of features that constitute a relatively high price. Here, besides features that are determined by the characteristics of a home, particularly (i) the canceling policy, (ii) the high quality preview picture, and (iii) the host’s verified profile contribute to an increased price. On the downside, the relatively high distance to the city center slightly contributes to a decrease in price. The force plot of prediction 1, on the other hand, exhibits the listing’s potential for improvement with respect to a higher price recommendation. This refers in particular to (i) using a high-quality preview picture (an MOS of 2.99 is rather low in our data) of (ii) another room (i.e., not the bedroom) and (iii) uploading more pictures.

\section{Conclusion}

In the present paper we developed and showcased a methodology that is tailored to guide the development and interpretation of ML-driven artifacts. To this end, we first positioned our paper according to existing research in DSR and predictive analytics in IS and presented available methodologies in these fields. Moreover, we elaborated on research in explainable AI and highlighted how the integration of corresponding knowledge into DSR may add value to IS research. We further uncovered the current lack of interpretable predictive modeling efforts in DSR and the need for a consistent methodology in this area by means of a respective literature review. Subsequently, we built our methodology upon the most important aspects of methodologies from DSR, predictive analytics, and explainable AI and provided valuable descriptions and execution guidelines. Finally, we showcased the methodology using the example of Airbnb price prediction and explanation for hosts that are new to the platform.

Throughout the paper, we were able to provide several contributions to research in IS. Among these are particularly the following: first, since our methodology is composed of knowledge from DSR and predictive analytics, we bridged the boundaries between these two fields. This is especially valuable for scholars who seek to develop prediction-oriented artifacts, as they are no longer compelled to decide whether to follow the DSR or predictive analytics paradigm—with the use of our methodology, they can do both at once. Moreover, our methodology is even applicable in prediction-oriented DSR studies that deliberately refrain from black box interpretation, as such efforts could simply follow the process through to evaluation and then disregard the understanding phase. Second, by considering the use of SHAP within the methodology, we introduced the proper use of a state-of-the-art approach to explainable AI from research in computer science to DSR. With this contribution, we particularly intend to pave the way for more studies in DSR that develop interpretable black box ML artifacts for prediction tasks. Third, in our use case, we showed how to extract insights from a black box ML model by the use of global explanations. Such extraction of insights may serve to stimulate theorizing and complement more traditional quantitative research in IS (i.e., explanatory modeling) in that it makes use of knowledge induction instead of deduction. In IS research, such inductive generation of knowledge is currently almost exclusively reserved for qualitative research in the sense of grounded theory \cite{Glaser1967,Muller2016}. However, most of the principles of grounded theory are transferrable to quantitative data and are well-suited to be applied along with predictive analytics \cite{Shmueli2011}. Hence, the use of our methodology may help to inform explanatory research efforts by enabling knowledge induction from quantitative data.

In addition to the contributions to research, our paper has certain practical implications. These are particularly the following: first, as our paper aims to encourage scholars to develop interpretable prediction-oriented ML artifacts, we contribute to the diffusion of ML in organizations, as such artifacts may serve as blueprints for real-world applications whose predictions can be explained to human decision-makers. The value of this contribution is twofold: (i) with the incorporation of explainable AI, ML applications can be employed in business cases that require system explainability, e.g., if decisions based on system predictions need to be justified legally or vis-à-vis prudential regulators; (ii) human decision-makers are more likely to adopt a system if they can comprehend the factors that constitute a prediction. Second, apart from researchers, practitioners may also use our methodology to develop artifacts—this might be particularly interesting within R\&D departments of organizations. Here, one promising avenue might be the incorporation of explainable AI within automated ML services, such as Google Cloud AutoML. Such services help consumers to develop ML-based applications—the incorporation of explainable AI might additionally support consumers in understanding an underlying model. Third, by showcasing the use of local explanations in our use case, we showed how explainable AI may assist practitioners in their decision-making. Thereby, we hope to inspire developers to build similar applications that may add value to practice.

Despite our contributions, our paper naturally comprises certain limitations, particularly with regard to our exemplary use case, which we deliberately kept compact to avoid exceeding an appropriate scope of this paper. One limitation arises from (i) the size of the Airbnb sample and (ii) its scope, that is, the sample could have been more comprehensive. This means that we might have trained a more generalizable model if we had used a sample with more observations, ideally from more regions in the USA and perhaps even other countries. Another limitation is our feature selection—although we already considered the use of numerous features, it would have been conceivable to incorporate more variables, such as points of interest near the individual homes (e.g., restaurants, gas stations, or supermarkets), additional information about neighborhoods (e.g., crime rates or weather data), or textual information from the descriptions of the homes. Moreover, by predicting average prices for each home, we did not take seasonal price fluctuations into account, which might have provided an even deeper understanding about pricing on Airbnb. However, since the focus of our use case was in showcasing our methodology, we consider its scope appropriate. In addition, the limitations may serve to encourage scholars to investigate the corresponding effects in future studies. Beyond the use case, we recognize that our paper puts a particularly strong emphasis on the use of SHAP and, thereby, neglects the application of other explainable AI approaches. Yet, as we pointed out throughout the paper, we currently find SHAP to be the most advanced approach to explainable AI. Therefore, we consider it well-suited to demonstrate the capabilities of explainable AI (i.e., in particular, the use of global/local explanations and feature interactions). Still, authors may use other approaches in the future, depending on the suitability with their projects or advances in explainable AI.

With our paper, we particularly intend to actively contribute to the development, acceptance, and use of ML-based applications, as we believe these hold great potential to facilitate knowledge induction in academia, increase productivity in organizations, and may assist us in our daily lives.

\bibliographystyle{unsrt}  
\bibliography{references}  

\newpage
\appendix

\section*{Appendix A: Feature Set}

\begin{longtable}{|l|l|}

\hline
\textbf{Feature Name}      & \textbf{Feature Description}                                                                                                                                                                                                                                                          \\ \hline
MOS                        & \begin{tabular}[c]{@{}l@{}}The mean opinion score is a measure of perceived image quality. \\ We applied it to homes’ preview images. MOS values are between \\ 1 (low quality) and 5 (high quality).\end{tabular}                                                                    \\ \hline
bathrooms                  & Number of bathrooms.                                                                                                                                                                                                                                                                  \\ \hline
bedrooms                   & Number of bedrooms.                                                                                                                                                                                                                                                                   \\ \hline
beds                       & Number of beds.                                                                                                                                                                                                                                                                       \\ \hline
person\_capacity           & Maximum number of guests.                                                                                                                                                                                                                                                             \\ \hline
picture\_count             & Number of uploaded pictures of a home.                                                                                                                                                                                                                                                \\ \hline
min\_nights                & Minimum number of overnight stays required for booking.                                                                                                                                                                                                                               \\ \hline
max\_nights                & Maximum number of overnight stays.                                                                                                                                                                                                                                                    \\ \hline
can\_instant\_book         & Boolean value that signalizes whether a home can be booked instantly.                                                                                                                                                                                                                 \\ \hline
verified\_card             & Boolean value that signalizes whether the host has a verified profile.                                                                                                                                                                                                                \\ \hline
en                         & Boolean value that signalizes whether the host speaks English.                                                                                                                                                                                                                        \\ \hline
preview\_wifi              & Boolean value that signalizes whether WiFi is a preview amenity.                                                                                                                                                                                                                      \\ \hline
preview\_kitchen           & Boolean value that signalizes whether a kitchen is a preview amenity.                                                                                                                                                                                                                 \\ \hline
preview\_heating           & Boolean value that signalizes whether heating is a preview amenity.                                                                                                                                                                                                                   \\ \hline
preview\_air\_conditioning & Boolean value that signalizes whether air conditioning is a preview amenity.                                                                                                                                                                                                          \\ \hline
preview\_free\_parking     & Boolean value that signalizes whether free parking is a preview amenity.                                                                                                                                                                                                              \\ \hline
preview\_washer            & Boolean value that signalizes whether a washer is a preview amenity.                                                                                                                                                                                                                  \\ \hline
preview\_pool              & Boolean value that signalizes whether a pool is a preview amenity.                                                                                                                                                                                                                    \\ \hline
preview\_hot\_tub          & Boolean value that signalizes whether a hot tub is a preview amenity.                                                                                                                                                                                                                 \\ \hline
CANCEL\_FLEXIBLE           & Boolean value that signalizes whether canceling is “flexible”.                                                                                                                                                                                                                        \\ \hline
CANCEL\_MODERATE           & Boolean value that signalizes whether canceling is “moderate”.                                                                                                                                                                                                                        \\ \hline
CANCEL\_STRICT\_14         & Boolean value that signalizes whether canceling is allowed within 14 days.                                                                                                                                                                                                            \\ \hline
CANCEL\_SUPER\_STRICT\_30  & Boolean value that signalizes whether canceling is allowed within 30 days.                                                                                                                                                                                                            \\ \hline
CANCEL\_SUPER\_STRICT\_60  & Boolean value that signalizes whether canceling is allowed within 60 days.                                                                                                                                                                                                            \\ \hline
picture\_bathroom          & Boolean value that signalizes whether the preview picture shows a bathroom.                                                                                                                                                                                                           \\ \hline
picture\_bedroom           & Boolean value that signalizes whether the preview picture shows a bedroom.                                                                                                                                                                                                            \\ \hline
picture\_dining\_room      & Boolean value that signalizes whether the preview picture shows a dining room.                                                                                                                                                                                                        \\ \hline
picture\_exterior          & Boolean value that signalizes whether the preview picture shows an exterior.                                                                                                                                                                                                          \\ \hline
picture\_interior          & Boolean value that signalizes whether the preview picture shows an interior.                                                                                                                                                                                                          \\ \hline
picture\_kitchen           & Boolean value that signalizes whether the preview picture shows a kitchen.                                                                                                                                                                                                            \\ \hline
picture\_living\_room      & Boolean value that signalizes whether the preview picture shows a living room.                                                                                                                                                                                                        \\ \hline
family\_amenities          & \begin{tabular}[c]{@{}l@{}}Boolean value that signalizes whether a home is equipped with amenities for families \\ (e.g., baby bath, babysitter recommendations, children’s dinnerware). \\ This feature results from hierarchical clustering of amenities.\end{tabular}              \\ \hline
basic\_amenities           & \begin{tabular}[c]{@{}l@{}}Boolean value that signalizes whether a home only comprises essential amenities \\ (e.g., WiFi, heating, smoke alarm). This feature results from hierarchical \\ clustering of amenities.\end{tabular}                                                     \\ \hline
normal\_amenities          & \begin{tabular}[c]{@{}l@{}}Boolean value that signalizes whether a home comprises more than only \\ essential amenities. This feature results from hierarchical clustering of amenities.\end{tabular}                                                                                 \\ \hline
accessible\_amenities      & \begin{tabular}[c]{@{}l@{}}Boolean value that signalizes whether a home is equipped with amenities that indicate \\ accessibility (e.g., no stairs and steps, wide entrance, extra space around bed). \\ This feature results from hierarchical clustering of amenities.\end{tabular} \\ \hline
entire\_home               & Boolean value that indicates whether a listing is an entire home.                                                                                                                                                                                                                     \\ \hline
private\_room              & Boolean value that indicates whether a listing is a private room.                                                                                                                                                                                                                     \\ \hline
hotel\_room                & Boolean value that indicates whether a listing is a hotel room.                                                                                                                                                                                                                       \\ \hline
shared\_room               & Boolean value that indicates whether a listing is a shared room.                                                                                                                                                                                                                      \\ \hline
other\_home                & \begin{tabular}[c]{@{}l@{}}Boolean value that indicates whether a listing is an alternative home \\ (e.g., houseboat, tree house, tiny house).\end{tabular}                                                                                                                           \\ \hline
sum\_foreign\_languages    & Number of foreign languages supported by the host.                                                                                                                                                                                                                                    \\ \hline
distance\_to\_center       & Distance from home to the city center, measured in kilometers.                                                                                                                                                                                                                        \\ \hline
city\_population           & Population of the city that corresponds to a home.                                                                                                                                                                                                                                    \\ \hline
population\_density        & Population density per square kilometer of the city that corresponds to a home.                                                                                                                                                                                                       \\ \hline
beds\_per\_bedroom         & Ratio of beds per bedroom.                                                                                                                                                                                                                                                            \\ \hline
persons\_per\_bedroom      & Ratio of persons per bedroom.                                                                                                                                                                                                                                                         \\ \hline
persons\_per\_bathroom     & Ratio of persons per bathroom.                                                                                                                                                                                                                                                        \\ \hline
\end{longtable}

\section*{Appendix B: Interaction Effects}
\label{app:B}

Here, we show a selection of conspicuous interaction effects from the XGBoost model using dependence plots. These are scatter plots that show values of a feature on the x-axis and the corresponding SHAP values on the y-axis. Moreover, the plotted dots are colored according to the values of an interacting feature.

Figure B1 depicts two plots, each showing the SHAP values of corresponding numbers of persons per bathroom. While the left plot colors dots according to the values of the binary feature “entire\_home”, the right colors these according to the values of “private\_room”. The figure unveils that negative SHAP values of “persons\_per\_bathroom” highly interact with private rooms, whereas entire homes heavily interact with SHAP values that are either positive or close to zero. This indicates that entire homes exert a positive impact on prices per night, while private rooms tend to be less expensive. Although this effect might seem trivial, it is highly suited to demonstrate the appearance of feature interaction in dependence plots.

\begin{figure}[H]
  \centering
  \includegraphics[width=1.0\textwidth]{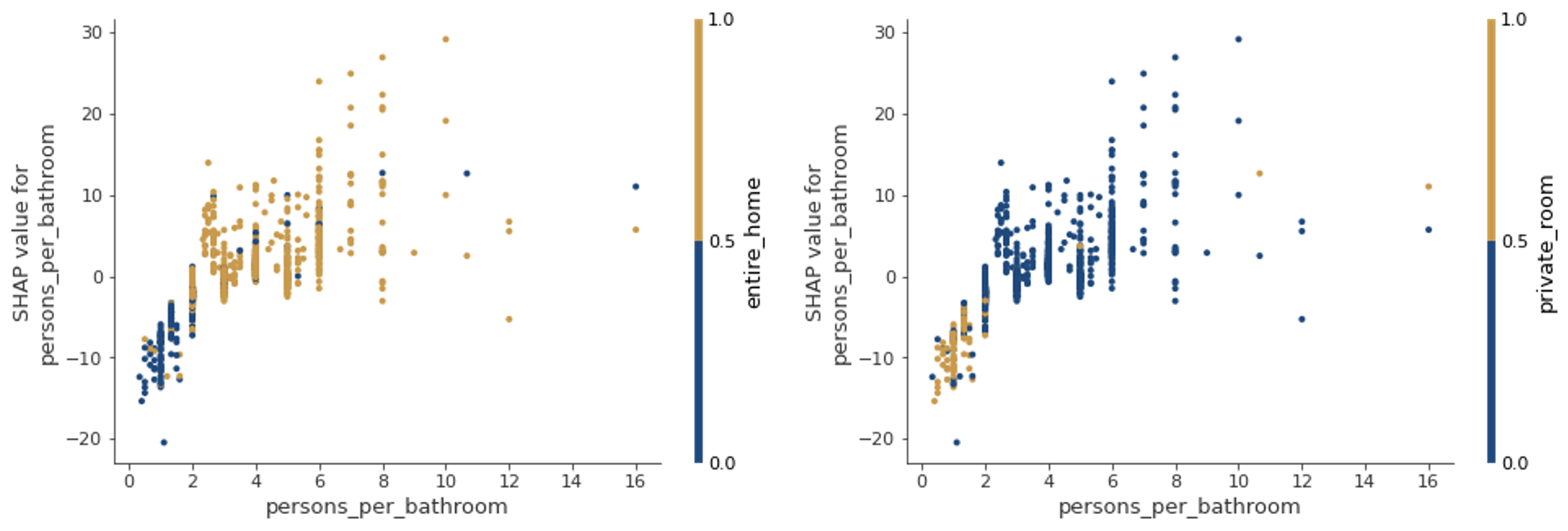}
  \caption*{Figure B1: Interaction effects between persons per bathroom and home types}
  \label{fig:fig11}
\end{figure}

Figure B2 depicts interaction effects between “population\_density” and (i) whether a home’s listing uses the amenity “heating” as a preview as well as (ii) whether the listing uses the amenity “free parking” as a preview. The figure’s left plot unveils homes in cities with a high population density to use heating more often as a preview compared to less densely populated cities. However, this does not mean that homes with heating as a preview amenity are more expensive because of heating—it is also conceivable that the increased SHAP values are due to rental prices that tend to be higher in bigger cities. It is also unclear if the value of “preview\_heating” is actually a result of higher population density or vice versa, as it might likewise be the case that other characteristics of the corresponding cities cause the values of both features. Moreover, the right plot of Figure B2 shows that free parking is not used as a preview amenity in cities with a particularly high population density. This effect is rather unsurprising, as we expect parking spots in highly populated cities to be rarer. Yet, having only the information from the figure’s right plot, we cannot confirm free parking to increase the SHAP values of low “population\_density” values considerably.

\begin{figure}[H]
  \centering
  \includegraphics[width=1.0\textwidth]{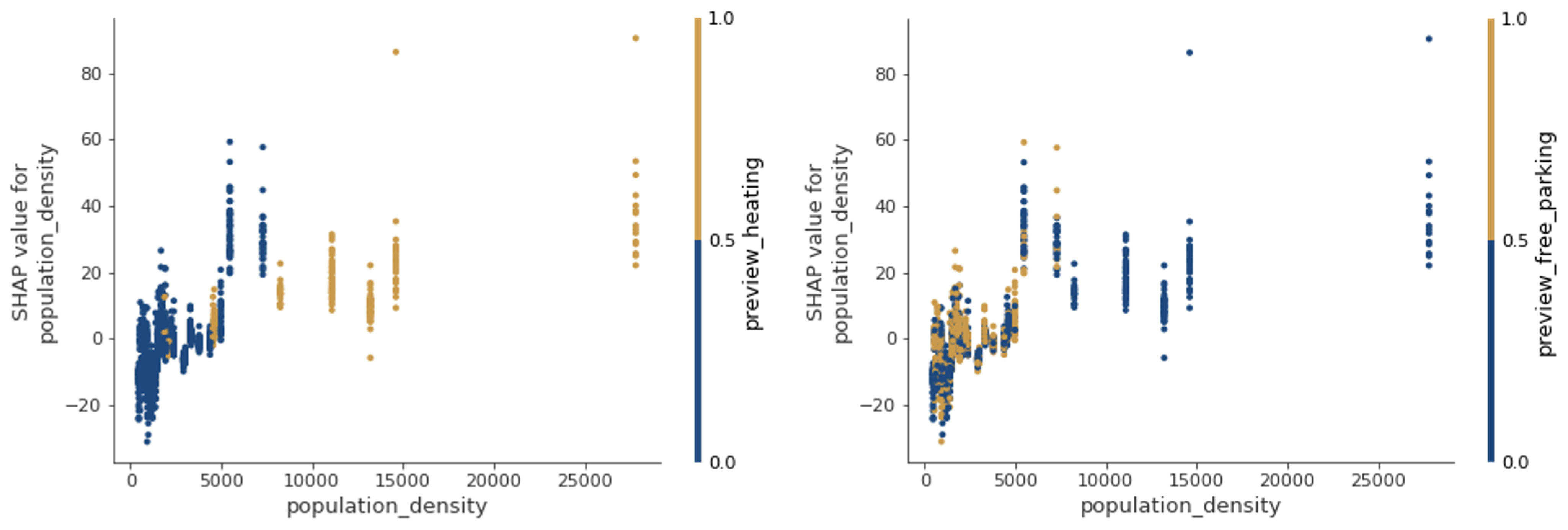}
  \caption*{Figure B2: Interaction effects between population density and preview amenities}
  \label{fig:fig12}
\end{figure}

Figure B3 depicts interaction effects of the binary variables “picture\_exterior” (i.e., whether the preview picture shows an exterior) and “picture\_bedroom” (i.e., whether the preview picture shows a bedroom). Here, the left plot unveils that for homes with an exterior picture, higher distances to city centers tend to contribute to an increase in a home’s SHAP values and vice versa. This might be due to the fact that hosts use attractive surroundings of secluded homes as preview pictures to compensate for their high distances to city centers—or, these homes are deliberately far from cities, because they aim to address guests who seek to stay in a place less urban. The plot in the middle of the figure further shows that, for these homes with an exterior picture, high population tends to increase prices. This leads to the following assumption: Homes with an attractive surrounding at the edge of a particularly large city seem to be more expensive. Finally, the right plot shows that, for homes with a bedroom as preview picture, higher population values interact with lower SHAP values and vice versa.

\begin{figure}[H]
  \centering
  \includegraphics[width=1.0\textwidth]{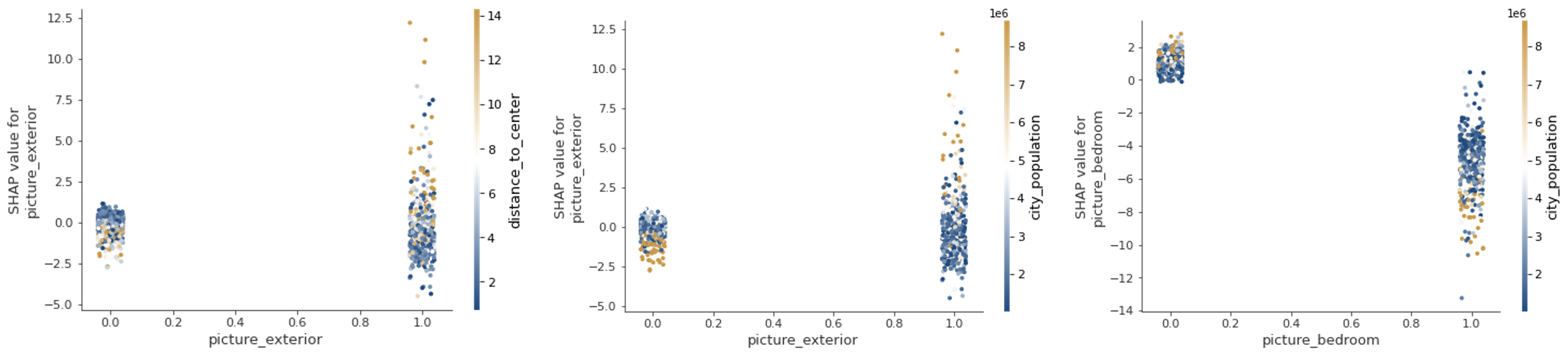}
  \caption*{Figure B3: Interaction effects between preview pictures and location-based features}
  \label{fig:fig13}
\end{figure}

Figure B4 depicts interactions of canceling policies. Interestingly, for homes with a flexible canceling strategy, more bedrooms tend to decrease SHAP values. Moreover, while the use of a pool as a preview amenity seems to increase SHAP values of homes with a moderate canceling strategy, we observe the opposite effect for homes with a strict canceling policy of 14 days.

\begin{figure}[H]
  \centering
  \includegraphics[width=1.0\textwidth]{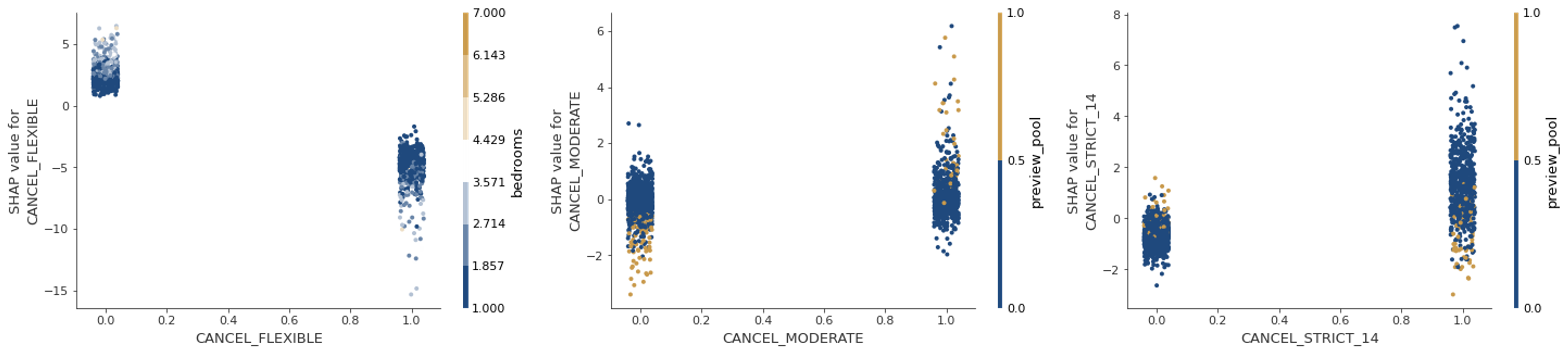}
  \caption*{Figure B4: Interaction effects of canceling policies}
  \label{fig:fig14}
\end{figure}

\end{document}